\documentclass{article}

\usepackage{PRIMEarxiv}

\usepackage{abstract}
\usepackage[utf8]{inputenc} 
\usepackage[T1]{fontenc}   
\usepackage{lmodern}       
\usepackage{url}            
\usepackage{booktabs}       
\usepackage{tabularx}       
\usepackage{amsfonts}       
\usepackage{amsmath}        
\usepackage{amsthm}         
\usepackage{nicefrac}       
\usepackage{microtype}      
\usepackage{placeins}
\usepackage{fancyhdr}       
\usepackage{graphicx}       
\graphicspath{{media/}}     
\usepackage[hidelinks]{hyperref}       

\newcolumntype{Y}{>{\centering\arraybackslash}X}

\hypersetup{
  pdftitle={Tensor-Network Formulation of the Traveling Salesman Problem and Variants},
  pdfauthor={Alejandro Mata Ali, Inigo Perez Delgado, Aitor Moreno Fdez.\ de Leceta},
  pdfkeywords={Traveling Salesman Problem, tensor networks, finite temperature heuristic, zero-temperature limit, Job Reassignment Problem}
}

\ifdefined\pdfgentounicode
  \IfFileExists{glyphtounicode.tex}{\input{glyphtounicode}}{}
\fi

\newtheorem{proposition}{Proposition}[section]
\newtheorem{lemma}{Lemma}[section]
\newtheorem{remark}{Remark}[section]
\theoremstyle{definition}
\newtheorem{algorithmblock}{Algorithm}
\theoremstyle{plain}

\title{\texorpdfstring{Tensor-Network Formulation of the Traveling Salesman Problem and Variants}{Tensor-Network Formulation of the Traveling Salesman Problem and Variants}
 }

\author{
  Alejandro Mata Ali \\
  Instituto Tecnol\'ogico de Castilla y Le\'on, Burgos, Spain\\
  \texttt{alejandro.mata.ali@gmail.com} \\
   \And
  I\~nigo Perez Delgado \\
  i3B Ibermatica, Parque Tecnol\'ogico de Bizkaia \\
  Ibaizabal Bidea, Edif. 501-A \\
  48160 Derio, Spain\\
  \texttt{iperezde@ayesa.com} \\
  \And
  Aitor Moreno Fdez.\ de Leceta \\
  Quantum Technologies and Systems Unit,\\
  LKS Next, MONDRAGON Corporation, Goiru 7,\\
  20500 Arrasate-Mondrag\'on, Gipuzkoa, Spain\\
  \texttt{aitormoreno@lksnext.com} \\
}

\begin{document}
\maketitle

\begin{abstract}
This work presents a tensor-network formulation of the Traveling Salesman Problem (TSP) and several of its variants. The approach represents candidate tours with tensor-network layers, weights them by Boltzmann factors, and enforces constraints through explicit counting filters. This formalism also yields an explicit tensor-network marginal formula whose zero-temperature, exact-arithmetic limit identifies an optimal feasible tour through a sequential marginal rule. At finite $\tau$ and finite precision, the implemented extraction is a heuristic whose behavior depends on numerical contrast, calibration, and near-degeneracies. We adapt the construction to several generalizations of the TSP and apply it to the Job Reassignment Problem, as a representative industrial integration. The experiments are deliberately small and illustrative; they contextualize the method against exact and heuristic references but do not establish general computational superiority over specialized classical solvers.
\end{abstract}


\section{Introduction}
The Traveling Salesman Problem (TSP)~\cite{flood_original_tsp,General_TSP,tsp_dynamic_hypercube} is a widely studied problem with great applicability for both basic research and industry, and is applied in route optimization, logistics, and scheduling. This problem asks for the shortest route in a graph that visits all its $N$ nodes once. It is an NP-hard problem, so no polynomial-time exact algorithm is known for the general case. Using brute force methods, the computational complexity of solving it is $\mathcal{O}\left(N!\right)$~\cite{flood_original_tsp}, and larger cases can be tackled using the branch-and-bound technique \cite{Branch_Bound,branch_bound_tsp_first}. Several exact algorithms exist to tackle this problem more efficiently, most notably the Held-Karp algorithm \cite{Held_TSP}, which solves it in $\mathcal{O}\left(N^2 2^N\right)$ and has exponential worst-case complexity. This makes finding certified optima for many industrial-scale cases impractical with general exact methods. However, in many applications, it is not necessary to obtain the optimal solution; a solution close to the optimum can be sufficient. This is because, often, the modeling of reality for the problem is not perfect, so the optimum of the model may not be the optimum in practice. In these cases, the time and cost required to calculate a certified optimum may not be justified relative to an approximate solution. To deal with this, approximate methods such as genetic algorithms \cite{Genetico} or heuristics \cite{Christofides} are applied, which output an approximate solution close enough to the optimal solution to be considered acceptable. In fact, these methods often output the optimal solution itself, although they do not offer any guarantee of it being such.

The emerging interest in quantum technologies due to their theoretical potential to solve some complex problems faster than classical algorithms \cite{Shor} has led to research and development of quantum algorithms for combinatorial optimization. This includes methods such as the Quantum Approximate Optimization Algorithm (QAOA) \cite{QAOA}, Variational Quantum Eigensolver (VQE) \cite{VQE} or Fixed-Point Grover Adaptive Search \cite{GroverAdaptive}, and also the particular case of TSP \cite{QTSP,GAS_TSP,Quantum_Speedups}. However, due to the current state of quantum hardware, in the Noisy Intermediate-Scale Quantum (NISQ) era, these algorithms cannot be implemented outside of simulators for many interesting instances.

This limitation has led to a growing interest in classical techniques that mimic the properties of quantum systems to perform calculations efficiently. Among them are Tensor Networks (TN) \cite{TensorNetwork,orus_tn_intro}, which are based on the properties of tensor operations to simulate quantum systems with restricted entanglement and support information compression techniques for quantum states \cite{MPS1,MPS2}. They can also apply operations not allowed in quantum systems, such as non-unitary operators, projections, or non-normalized states. Although some of these operations, such as imaginary time evolution, can be performed with known techniques \cite{Evolution}, linking several of these operations is challenging, especially with projection, while maintaining low storage requirements. TN techniques have been used on several occasions for combinatorial optimization problems \cite{TTOpt,GEO,Combin}, while formal exact-arithmetic tensor-network formulations tailored to TSP using the explicit counting-filter and iterative marginal-contraction scheme developed here remain, to the best of our knowledge, underexplored. Recent tensor-network generator approaches have also been explored for TSP using MPS/Born-machine models and autoregressive sampling~\cite{tn_geo_tsp}; this line is complementary to the full counting-filter contraction framework considered here. Although these methods are designed for a large number of cases, if they are not specifically adjusted to the particular problem, they require excessive bond dimensions or TN contraction times.

This work presents a tensor-network construction for representing exact-arithmetic marginal formulas for the TSP and several of its generalizations. The formulation represents all feasible and infeasible assignments in a tensor product space, modifies the amplitudes of the combinations with their costs, applies the constraints of the problem, and finally extracts the highest-weight feasible state in the zero-temperature limit. One of the main uses of this construction is the possibility of obtaining approximate solutions by heuristically eliminating layers to reduce computational complexity and memory cost, allowing us to address a larger number of cases or to couple it to other future algorithms. We do not claim a general computational advantage for TSP; the exact rule is exponential, and the finite-$\tau$ implementation should be read as a calibrated finite-precision heuristic.

The main contributions of this work are:
\begin{itemize}
    \item The explicit tensor-network marginal equation whose zero-temperature, exact-arithmetic limit identifies an optimal feasible solution for the TSP and its generalizations.
    \item The definition of the matrix product operator layers for applying counting constraints, which is useful for several other problems.
    \item  A sequential marginal-extraction rule for the zero-temperature exact-arithmetic solution of the TSP and its generalizations, together with a finite-$\tau$ heuristic approximation for larger instances.
    \item An analysis of a finite-precision implementation of this  construction for the TSP and for the Job Reassignment Problem (JRP).
\end{itemize}

This work is structured as follows. First, Sec.~\ref{sec:background} introduces the problem definition and a brief background of the state-of-the-art algorithms for solving it. Second, Sec.~\ref{sec:tn_tsp} introduces the TN formulation and marginal equation for the TSP. Then, Sec.~\ref{sec:general_tsp} presents several generalizations of the TSP and the corresponding TN construction. Finally, Sec.~\ref{sec:experiments} presents small illustrative experiments with the finite-$\tau$ implementation for the original TSP and the ONCE Job Reassignment proof of concept.

All code and implementations of this work are available in the GitHub repository \url{https://github.com/DOKOS-TAYOS/Traveling_Salesman_Problem_with_Tensor_Networks}.

\section{Background}\label{sec:background}
\begin{figure}[h]
  \centering
  \includegraphics[width=0.7\linewidth]{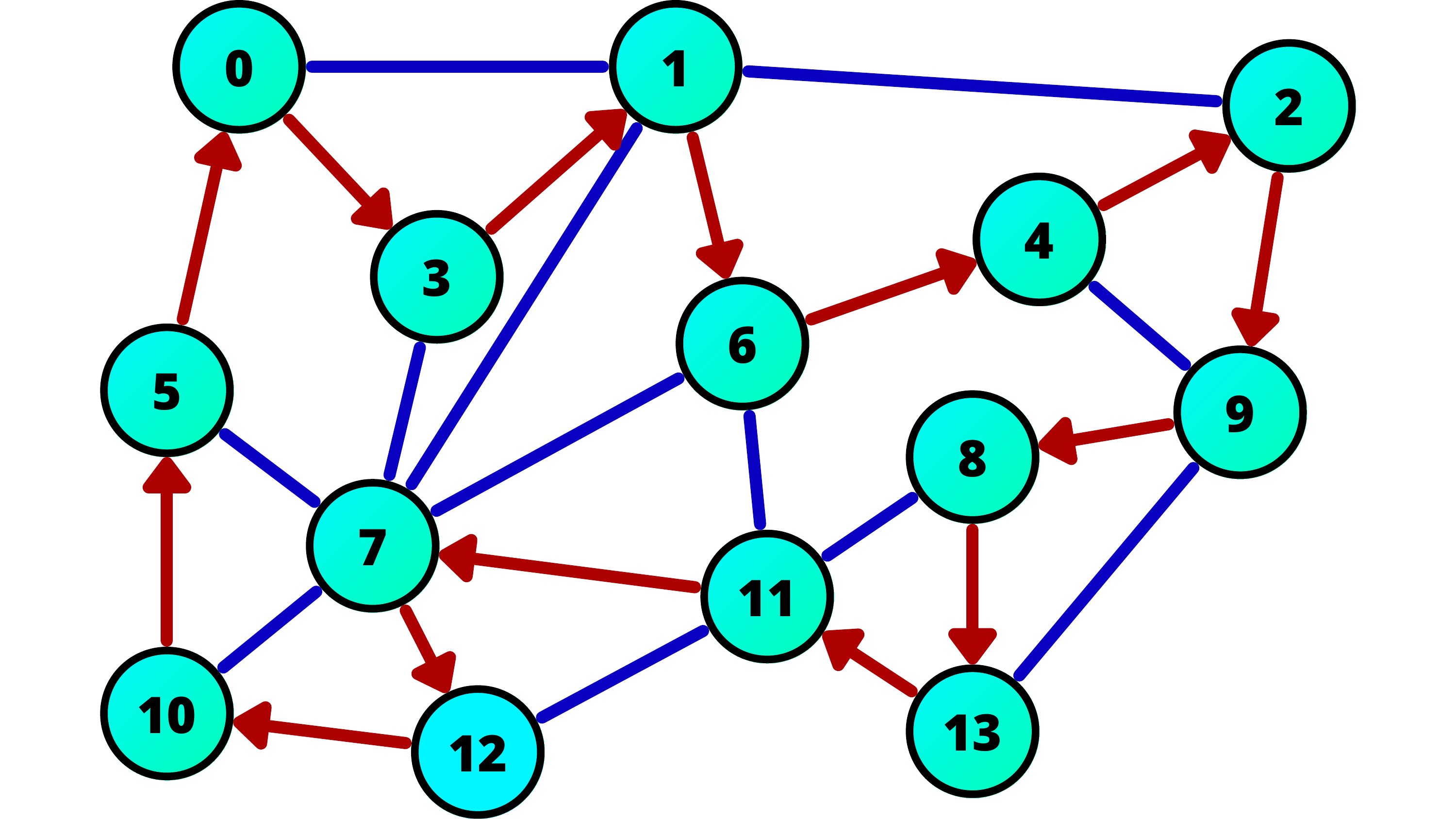}
  \caption{TSP graph with the solution $\vec{x} = (0,3,1,6,4,2,9,8,13,11,7,12,10,5)$. The blue edges represent possible paths that were not taken, and arrowed red edges represent the path taken by the solution. Edges with weight considered infinite are not represented.}
  \label{fig:TSP_Nodes}
\end{figure}

Given a fully connected graph $\mathcal{G} = (V,E)$ of $N=|V|$ nodes, with a cost $E_{i,j}$ associated with each edge that connects the $i$-th node with the $j$-th node, the objective is to determine the route that traverses the edges with the lowest possible total cost, visits each node exactly once and then return to the starting one (return condition). The graph can be directed or undirected, and the edges can be costs associated to the direction of travel. The solutions can be expressed as a vector $\vec{x}$ of integers, having its components $x_t$ the node visited at step $t$. An example of the solution can be seen in Fig.~\ref{fig:TSP_Nodes}.

With this formulation, the problem can be expressed as
\begin{gather}
    \vec{x}_{opt} = \arg\min_{\vec{x}} \left(C\left(\vec{x}\right)\right)\nonumber\\
    x_t\in[0, N-1]\ \forall t\in [0, N-1]\nonumber\\
    x_t \neq x_{t'} \qquad \forall t,t' \in [0, N-1]| t\neq t'\nonumber \\
    C(\vec{x}) = \sum_{t=0}^{N-1} E_{x_t, x_{t+1}}, \label{eq:form_TSP_1}
\end{gather}
where $C(\vec{x})$ is the cost function of the problem, $E_{i,j}$ is the cost of the travel from node $i$ to node $j$ and $x_N=x_0$ is the starting and final node. If two nodes $i$ and $j$ are not meant to be connected, $E_{i,j}=\infty$. The tensor that stores $E_{i,j}$ includes the information of whether it is a directed or undirected graph, which for the TN method is indifferent in terms of complexity. In the non-symmetrical cost case, $E_{i,j}\neq E_{j,i}$. The problem can be naturally extended for the $x_0 \neq x_N$ just by adjusting the summation limits.

The problem can also be generalized to the Time Dependent TSP (TDTSP)~\cite{original_tdtsp} where, at each step $t$, the costs of going from node $i$ to node $j$ change and the cost function can be generalized to 
\begin{equation}
    C(\vec{x}) = \sum_{t=0}^{N-1} E_{t,x_t, x_{t+1}},    
\end{equation}
where $E_{t,i,j}$ is the cost of moving from node $i$ to node $j$ at step $t$. If there is no return condition, $E_{N-1,a,b}=0\ \forall a,b$.

An extra cost can be included for arriving at a node at each time step. This would represent, for the canonical interpretation of the TSP, the time spent by the salesman actually making each sale at node $a$. This would be an extra linear term, so the cost would be
\begin{equation}
    C(\vec{x}) = \sum_{t=0}^{N-1}\left(E^0_{t,x_t}+ E_{t,x_t, x_{t+1}}\right),\label{eq:gen_TSP_QUDO}
\end{equation}
with $E^0_{t,i}$ as the cost of being at node $i$ at step $t$. Fixed or forbidden visits are better represented by local projectors rather than by $-\infty$ energies. To force $x_t=a$, we use the local projector
\begin{equation}
    P_t(i)=\mathbf{1}_{\{i=a\}},
\end{equation}
while forbidding $x_t=a$ is implemented with
\begin{equation}
    P_t(i)=\mathbf{1}_{\{i\neq a\}}.
\end{equation}
These projectors multiply the local tensor or initialization layer at time $t$. A forbidden assignment may be formally described by a $+\infty$ penalty, which is equivalent to setting the corresponding tensor entry to zero, whereas $-\infty$ should only be understood as a limiting shorthand and is not used as a finite tensor entry.

For a closed route, the linear term can be absorbed into the quadratic term by defining
\begin{equation}
    \hat{E}_{t,i,j}=E_{t,i,j}+E^0_{t,i},
\end{equation}
so that
\begin{equation}
    C(\vec{x}) = \sum_{t=0}^{N-1}\hat{E}_{t,x_t, x_{t+1}}.\label{eq:general_TSP}
\end{equation}
For an open path, the terminal linear contribution can be left as a separate local term or absorbed by introducing an auxiliary terminal state.

This problem can be expressed as a Tensor Quadratic Unconstrained Discrete Optimization (T-QUDO) problem with nearest-neighbor interactions in a one-dimensional chain with the additional constraint of non-repetition. In the continuous case, the cost function can be expressed as
\begin{equation}
    C(\vec{x}) = \sum_{t=0}^{N-1}\hat{E}(T_t(\vec{x}),x_t, x_{t+1}),\quad T_t(\vec{x})=\sum_{t'=0}^{t-1}\hat{E}(T_{t'}(\vec{x}),x_{t'}, x_{t'+1}),
\end{equation}
where the real arrival time $T_t$ at step $t$ changes the cost of the travel from $x_t$ to $x_{t+1}$.

There are several algorithms in the state-of-the-art to solve this. A classical reference is the Held-Karp algorithm~\cite{Held_TSP}, which solves it in $\mathcal{O}\left(N^2 2^N\right)$ using dynamic programming and is a standard exact dynamic-programming baseline. This algorithm obtains the optimum in exact arithmetic, but its complexity scales exponentially, limiting its application to practical instances. Other perspectives are those proposed by branch-and-bound algorithms.

Branch-and-Bound explores a search tree whose nodes encode partial decisions about the tour (typically edges forced in or out). From a root relaxation, the algorithm recursively branches on an undecided edge $\{i,j\}$, creating two subproblems: one that includes $\{i,j\}$ and one that forbids it. Each node $i$ is equipped with a computable lower bound $C(i)$ on the cost of any tour consistent with the decisions at $i$. Common bounds come from relaxations such as the assignment problem / reduction of cost-matrix~\cite{branch_bound_tsp_first} and the minimum $1$-tree~\cite{spanning_tree} obtained by a spanning tree on $V\setminus\{r\}$ plus two edges incident to a root $r$, possibly strengthened by Lagrangian penalties. Let $z^\star$ denote the best feasible tour found so far (the incumbent). Any node with $C(i)\ge z^\star$ is pruned. Otherwise, if $i$ is integral (i.e., its relaxation yields a Hamiltonian cycle), $z^\star$ is updated; if not, another branch is created. The best-first search (expanding the node with the smallest bound) is standard, though depth-first with bounding is also used. Cutting planes~\cite{large_tsp} (e.g., subtour elimination constraints) can be added to tighten the relaxation before branching, yielding a branch-and-cut scheme~\cite{Branch_Bound}. Even when its worst-case scenario is also $\mathcal{O}(N!)$, this method is more practically applicable than the Held-Karp one.

There exist several approximate algorithms which are useful in practical instances. These heuristics are fast and useful for generating initial solutions for local search algorithms. The first example is the Christofides algorithm~\cite{Christofides} for metric spaces, which uses the shortest spanning tree of the graph to approximate the solution. Its computational complexity is $\mathcal{O}(N^3)$ and is guaranteed that the cost of its solution is at most $3/2$ of the optimal solution cost.

Another possibility is constructive heuristics, which generate a tour from scratch, incrementally adding cities based on simple rules. The Nearest Neighbor (NN) heuristic is among the simplest, always choosing the closest unvisited city~\cite{heuristic_tsp,johnson1997traveling}. Despite its efficiency, NN often results in suboptimal tours. Variants like the Cheapest Insertion and the Clarke-Wright savings heuristic improve performance by considering cost-effective ways to insert cities~\cite{clarke_tsp,johnson1997traveling}. Local search methods start with an initial tour and iteratively improve it by exchanging segments. The \textit{2-opt} heuristic removes two edges and reconnects the tour in a different way if the new configuration shortens the route~\cite{2-opt}. \textit{3-opt} generalizes this considering three edge exchanges. The Lin-Kernighan (LK) algorithm dynamically adapts the number of edges exchanged and is widely regarded as one of the most powerful local search heuristics for TSP~\cite{lin_tsp}. An enhanced implementation, LKH by Helsgaun, incorporates advanced features such as candidate sets and $k$-opt moves, and is known to solve large instances with near-optimal solutions~\cite{lin-ken-tsp-large,Helsgaun2009}.

Genetic Algorithms are population-based metaheuristics that evolve solutions over generations using selection, crossover, and mutation. For the TSP, crossover operators that respect the permutation structure are essential. The Edge Assembly Crossover (EAX) proposed by Nagata and Kobayashi is one of the most effective for TSP~\cite{nagata_genetic}. Their GA-EAX algorithm consistently finds optimal or near-optimal tours, often outperforming other metaheuristics in large instances. Several other metaheuristics have been applied to TSP, including Simulated Annealing (SA)~\cite{simulated_annealing}, and Tabu Search~\cite{tabu_search}. These methods explore the solution space by accepting worse solutions under certain conditions to escape local minima. Although often less effective than LK-based methods or GA-EAX in terms of solution quality, they are valuable for their simplicity and robustness in various settings. Hybrid algorithms that combine these approaches with local search have demonstrated improved results.

Other variations, such as the TDTSP, have been studied using linear programming~\cite{original_tdtsp}, branch-and-cut-and-price algorithms~\cite{time_dependent_tsp} and heuristics~\cite{tdtsp_heuristic}.

\FloatBarrier
\section{\texorpdfstring{Tensor-network formulation for the TSP}{Tensor-network formulation for the TSP}}\label{sec:tn_tsp}
This section presents a tensor-network formulation for the case of the TDTSP described in the previous section, so that in each iteration the node corresponding to its time step on the trajectory is obtained. The construction consists of a logical tensor network that takes into account all the possible tours, reduces the elements associated with each one by its cost, and finally removes the incompatible tours. This procedure defines marginal weights whose zero-temperature, exact-arithmetic limit selects an optimal value for each variable by contracting its corresponding tensor network. All the process is encoded in a represented tensor, which is the result of contracting the tensor network. This tensor consists of two pieces: the indices and the elements. The indices of this represented tensor are associated with the value of its corresponding variable. For example, the indices of the tensor $T_{i,j,k}$ represent the possible values for the variables $i,j,k$. The elements of the tensor are associated with fixed values of its indices. For example, $T_{0,3,2}$ is associated with the index values $i=0,j=3,k=2$. Due to the fact that a tensor contains elements associated with all possible values of its indices, the tensor is able to have a `superposition' of all possible combinations of variable values, associated to a value. So each combination has an associated value, in this work called \textit{amplitude} in analogy with the quantum systems, which can be manipulated to obtain certain results. At first sight, this can be confused with brute force search, due to the management of all possible combinations. However, the distinction from it is obtained by the tensor network formalism. The tensor network that represents that tensor is constructed by pieces, but its computation is not performed in the same way. Instead of explicitly computing the tensor $T$ in the order in which the pieces are designed, a reduced version is constructed, with the desired minimal properties needed to solve the problem and then computed along a contraction direction to reduce the storage requirements and computation time.

First, Sec.~\ref{ssec:construction} presents the construction of the tensor network to obtain a variable value. Second, Sec.~\ref{ssec:contraction} indicates how to perform the iterative method and the contraction of the tensor network to obtain the value of each variable. Then, Sec.~\ref{ssec:complexity} analyzes the computational complexity of the algorithm. Finally, Sec.~\ref{ssec:approximation} presents an approximate version of the algorithm to reduce its computational complexity.

\subsection{Tensor network construction}\label{ssec:construction}
This method evaluates a uniform superposition of all possible values of $\hat{N}$ variables, one for each timestep, in a qudit formalism that can each be in $\hat{N}$ different basis states, one for each node $i$. The number $\hat{N}$ will vary according to the instance to solve. For the formulation in Eq.~\eqref{eq:form_TSP_1}, the closed stationary route is invariant under cyclic shifts, so one node can be fixed without loss of generality. For the time-dependent TSP, the same start-fixing reduction is valid only if the time origin is also arbitrary or cyclic; otherwise the start must be treated as fixed input, or all possible starts must be considered. Under the stationary closed-route convention, it is fixed $x_{N-1}=N-1$. That is, the node visited in the last timestep is the last node of the graph. Therefore, it is only necessary to solve the problem for the other $\hat{N}=N-1$ steps of the route with $\hat{N}=N-1$ nodes. If there is no return condition, it should be solved with all the $\hat{N}=N$ variables. If the route must be between two fixed points, $x_0$ and $x_{N-1}$ are fixed, and the algorithm is performed on the other $\hat{N}=N-2$.

In this superposition, an imaginary time evolution is applied to decrease the amplitudes of the combination $\vec{y}$ exponentially with the cost $C(\vec{y})$ of the combination, based on the methods presented in \cite{Tridiag}. After that, a set of projectors is applied to preserve only the subspace that satisfies the constraints, with the inclusion of new layers of Matrix Product Operator (MPO) based on the methods presented in \cite{Combin}, extended to cases of non-binary variables and transmission of signals over long distances in the tensor network. Finally, the remaining state of maximum amplitude is determined with an iterative partial trace. The first part will be responsible for minimizing the cost function \eqref{eq:form_TSP_1} for the unconstrained nearest-neighbor T-QUDO problem, while the second part will be responsible for the fulfillment of the constraints.

\begin{figure}[h]
  \centering
  \includegraphics[width=\linewidth]{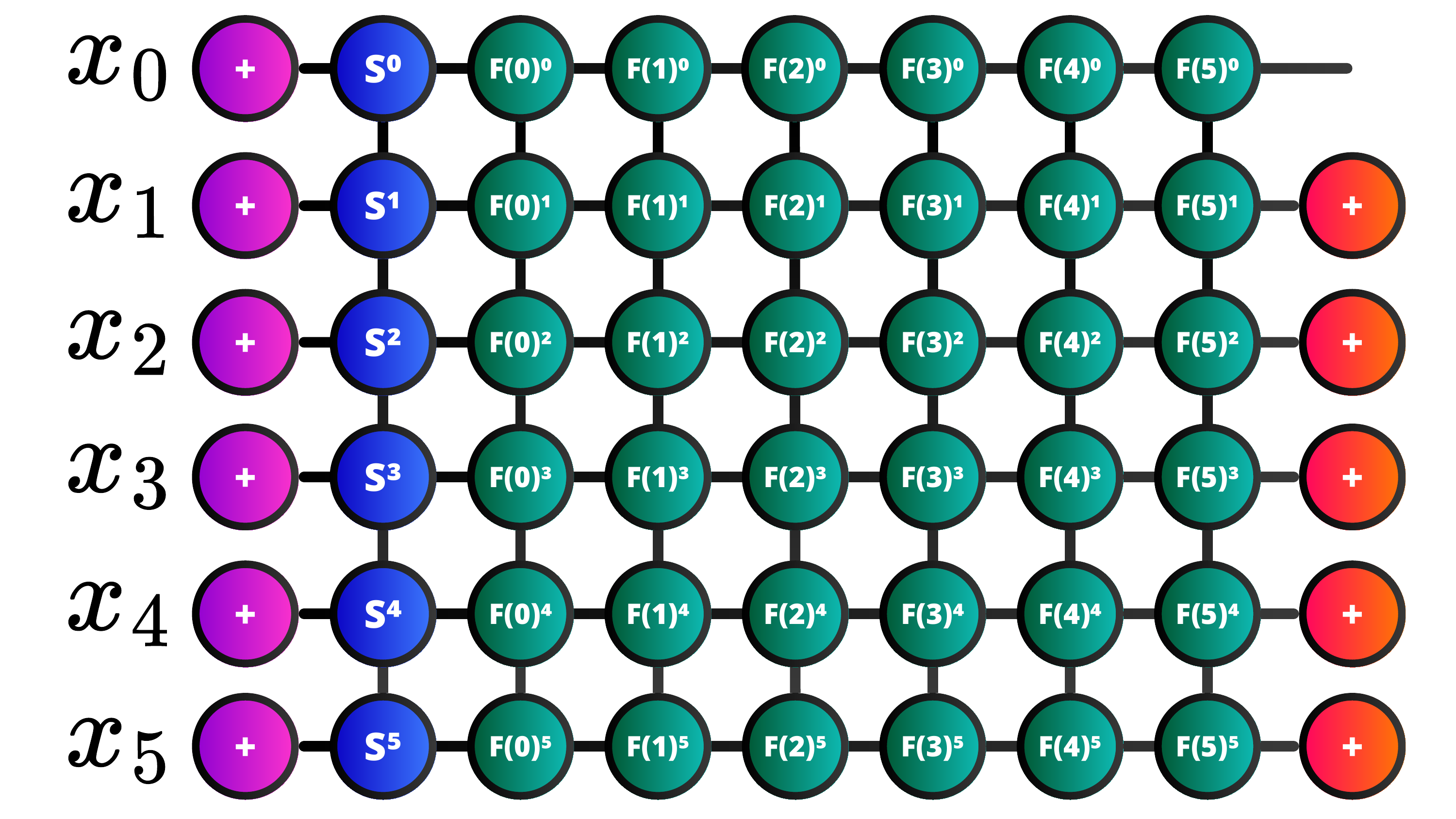}
  \caption{TSP tensor network, with the initialization and tracing `+' tensor layer, the `$S$' optimization layer and the `$F$' filter layers. For visual clarity, the sketch shows the full family of node filters; in the full constraint encoding one of them can be omitted because $\hat{N}-1$ exact-once filters already force the remaining node by counting.}
  \label{fig:TensorNetworkTSP}
\end{figure}

The layers consist of the initialization layer `+' and the optimization layer `$S$', the ones introduced by the work \cite{Tridiag}, $\hat{N}-1$ filter layers `$F$', introduced in this work, and the trace layer `+', also introduced by \cite{Tridiag}. The overall initialization/evolution/projection/trace viewpoint is also consistent with the broader MeLoCoToN formalism \cite{MeLoCoToN}. These first two layers are extensively explained in \cite{Tridiag}, so only a brief explanation of the result of them is given.

The first layer generates a state of maximum superposition of all possible combinations of the subindex values, a variable corresponding to each subindex.
\begin{equation}
    \psi^0_{\vec{y}}=1 \ \forall \vec{y}.
\end{equation}

The second layer performs the evolution in imaginary time as a diagonal operator
\begin{equation}
    U_{\vec{y},\vec{z}}= e^{-\tau C(\vec{y})}\delta_{\vec{y},\vec{z}}
\end{equation}
$\delta_{\vec{a},\vec{b}}$ being the Kronecker delta, whose value is $1$ if $\vec{a}$ and $\vec{b}$ are the same and $0$ otherwise. This operator associates with each combination an amplitude proportional to its cost with a damping factor $\tau$ for scaling, so that higher-cost configurations receive smaller amplitudes. Thus, the tensor after applying this layer is
\begin{equation}
    \psi^1_{\vec{y}}=e^{-\tau C(\vec{y})} \ \forall \vec{y}.
\end{equation}

After that, constraints are imposed with the inclusion of new MPO filter layers `$F$'. Since the constraint is non-repetition, the $R$ operator represented by the set of tensor layers `$F$' can be seen as a squared Levi-Civita symbol, so that it is $0$ if any index is repeated and $1$ if they are all different
\begin{equation}
    R_{\vec{y},\vec{z}} =  |\epsilon_{y_0,y_1, \dots,y_{\hat{N}-1}}|^2\delta_{\vec{y},\vec{z}}
\end{equation}
After applying this layer, we only have the combinations that satisfy the constraints, each with its corresponding amplitude. The state is
\begin{equation}
    \psi^2_{\vec{y}}=|\epsilon_{y_0,y_1, \dots,y_{\hat{N}-1}}|^2e^{-\tau C(\vec{y})} \ \forall \vec{y},
\end{equation}
which is equivalent to
\begin{equation}
    \psi^2_{\vec{y}}=e^{-\tau C(\vec{y})} \ \forall \vec{y}\in\mathcal{H},
\end{equation}
$\mathcal{H}$ being all the combinations which do not repeat the values of two variables.

Finally, to extract the solution, the trace layer `+' is added, also from \cite{Tridiag}. The final tensor after tracing is
\begin{equation}
    P^{0}_{y_0} = \sum_{y_1,y_2,\dots}|\epsilon_{y_0,y_1, \dots,y_{\hat{N}-1}}|^2 e^{-\tau C(\vec{y})},
\end{equation}
a vector whose component $i$ sums up all the amplitudes of the combinations whose first component is $y_0$. 

For the empty prefix $p=\emptyset$, the components of $P^0$ are precisely the partial partition sums
\begin{equation}
    P^0_a = Z_{\emptyset}(a;\tau),
\end{equation}
where $Z_p(a;\tau)$ is defined below for a generic prefix.

\begin{proposition}\label{prop:boltzmann_marginals}
Let $p=(x_0,\dots,x_{r-1})$ be a fixed prefix and let $\mathcal{F}_p$ be the set of feasible routes extending $p$. For each candidate $a$ at position $r$, define
\begin{equation}
    \mathcal{F}_{p,a}=\{x\in\mathcal{F}_p:x_r=a\},
\end{equation}
and
\begin{equation}
    Z_p(a;\tau)=\sum_{x\in\mathcal{F}_{p,a}}e^{-\tau C(x)}.
\end{equation}
Also define
\begin{equation}
    C_p^\star=\min_{x\in\mathcal{F}_p}C(x),
\end{equation}
and let $I_p^\star$ be the set of candidates $a$ such that $\mathcal{F}_{p,a}$ contains at least one optimal continuation of cost $C_p^\star$. For $a\notin I_p^\star$ with $\mathcal{F}_{p,a}\neq\emptyset$, define
\begin{equation}
    \Delta_a=\min_{x\in\mathcal{F}_{p,a}}C(x)-C_p^\star>0.
\end{equation}
Then, for every $b\in I_p^\star$,
\begin{equation}
    Z_p(b;\tau)\ge e^{-\tau C_p^\star},
\end{equation}
while for every $a\notin I_p^\star$,
\begin{equation}
    Z_p(a;\tau)\le |\mathcal{F}_{p,a}|e^{-\tau(C_p^\star+\Delta_a)}.
\end{equation}
Consequently,
\begin{equation}
    \frac{Z_p(a;\tau)}{Z_p(b;\tau)}\le |\mathcal{F}_{p,a}|e^{-\tau\Delta_a}.
\end{equation}
Hence, in the limit $\tau\to\infty$, the marginal weight of every suboptimal branch is exponentially suppressed relative to any branch that can be extended to a global optimum. For finite $\tau$, the same dominance holds whenever $\tau$ is sufficiently large compared with the gaps $\Delta_a$ and the number of feasible continuations.
\end{proposition}

\begin{proof}
If $b\in I_p^\star$, then by definition there exists $x^\star\in\mathcal{F}_{p,b}$ with $C(x^\star)=C_p^\star$, so the corresponding term already gives $Z_p(b;\tau)\ge e^{-\tau C_p^\star}$. If $a\notin I_p^\star$ and $\mathcal{F}_{p,a}\neq\emptyset$, every element of $\mathcal{F}_{p,a}$ has cost at least $C_p^\star+\Delta_a$, hence
\begin{equation}
    Z_p(a;\tau)=\sum_{x\in\mathcal{F}_{p,a}}e^{-\tau C(x)}\le \sum_{x\in\mathcal{F}_{p,a}}e^{-\tau(C_p^\star+\Delta_a)}=|\mathcal{F}_{p,a}|e^{-\tau(C_p^\star+\Delta_a)}.
\end{equation}
Dividing by the lower bound for $Z_p(b;\tau)$ yields the ratio estimate.
\end{proof}

\begin{remark}
For finite $\tau$ and finite-precision arithmetic, the extraction quality depends on the separation between competing costs and on the number of feasible continuations. In practice, rescaling and log-sum-exp type evaluations can be useful to avoid underflow or loss of contrast between nearly degenerate branches.
\end{remark}

Fig.~\ref{fig:TensorNetworkTSP} shows the total tensor network that will be built after connecting the tensors of the layers and the layers to each other.

We have the + tensors that implement the uniform superposition of all possible combinations as our qudits, the $S$ tensor layer implements the imaginary time evolution using the $\hat{E}_{i,x_i, x_{i+1}}$ tensor of our problem, and each $F(a)$ tensor layer enforces that node $a$ appears exactly once in the combination. For any tensor $T^i$, the superscript $i$ indicates that the tensor is connected to the qudit line $i$.

For clarity, the cost layer can be written explicitly in terms of the physical index $i$ for the current candidate node, the copied physical index $j$, the incoming bond index $k$ that carries the previous node, and the outgoing bond index $l$ that carries the current node to the next site. For a closed route with fixed boundary node $x_{\mathrm{start}}$, one convenient convention is
\begin{equation}
    S^0_{i j l}
    =
    \delta_{ij}\mathbf{1}_{\{l=i\}}
    e^{-\tau \hat{E}_{0,x_{\mathrm{start}},i}},
\end{equation}
\begin{equation}
    S^t_{i j k l}
    =
    \delta_{ij}\mathbf{1}_{\{l=i\}}
    e^{-\tau \hat{E}_{t,k,i}},
    \qquad 1\le t\le \hat{N}-2,
\end{equation}
and, at the last free position,
\begin{equation}
    S^{\hat{N}-1}_{i j k}
    =
    \delta_{ij}
    e^{-\tau\left(\hat{E}_{\hat{N}-1,k,i}+\hat{E}_{\mathrm{return},i,x_{\mathrm{start}}}\right)}.
\end{equation}
Thus the incoming node $k$ is used only to evaluate the edge entering the current candidate $i$, the factor $e^{-\tau \hat{E}}$ supplies the Boltzmann weight, and the outgoing bond propagates the current node to the next cost tensor. For open paths or different fixed endpoints, the two boundary factors are replaced by the corresponding fixed predecessor or successor terms.

The dimension of the indices of the tensors `+' and `$S$' will be in all cases $\hat{N}$, to be able to indicate which is the state of that qudit and to communicate it to the adjacent qudit. The `$F$' tensors will have indices of such dimensions that they can receive and send the $\hat{N}$ possible states of their qudit, but the signal they send to the adjacent `$F$' tensor of the same layer will be binary.

\begin{figure}[h]
  \centering
  \includegraphics[width=\linewidth]{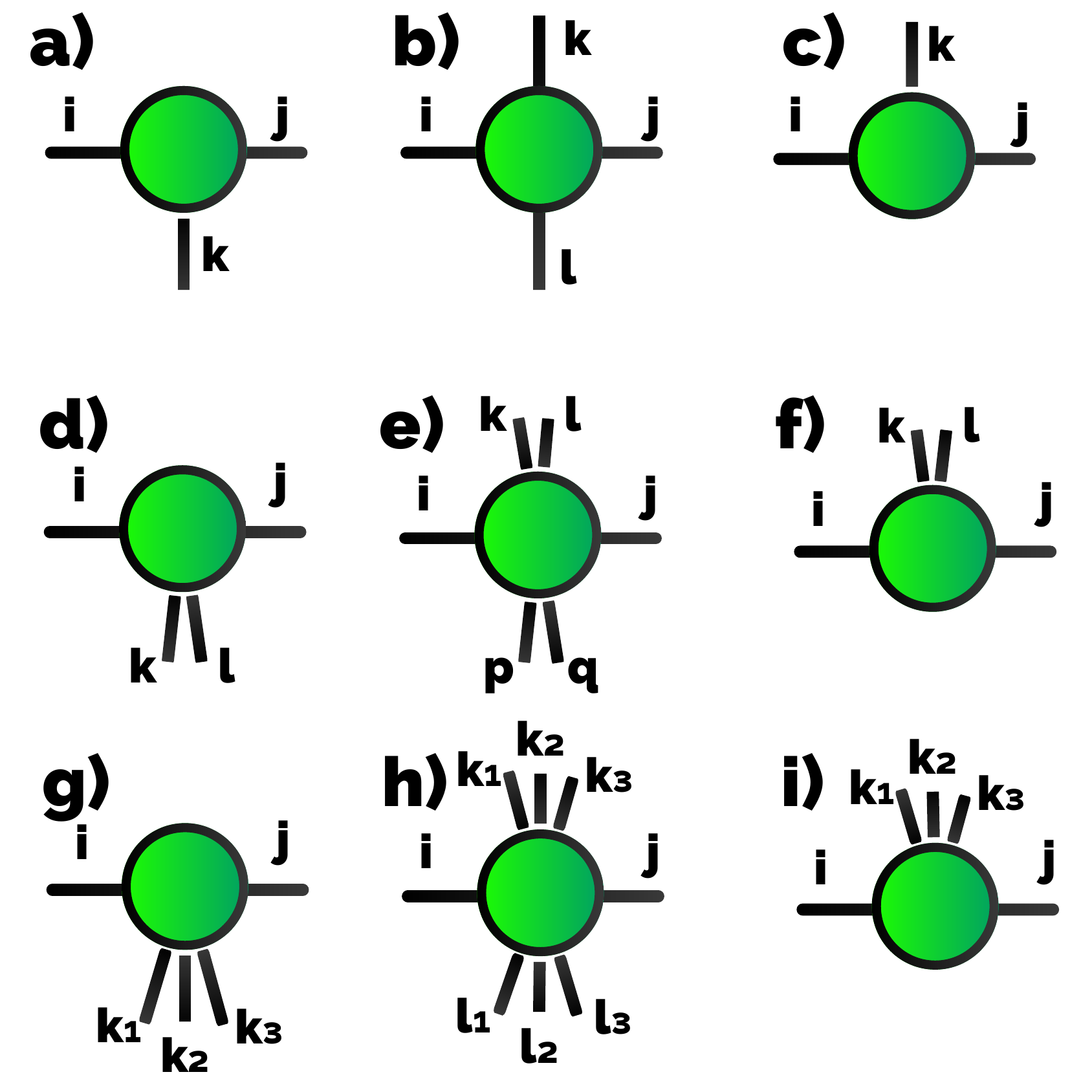}
  \caption{Notation for all indices used in all tensors in this paper. a), b), c) Tensors with one bond index, d), e), f) tensors with two bond indices, g), h), i) tensors with three bond indices.}
  \label{fig:TSP_Tensors}
\end{figure}

Using the notation presented in Fig.~\ref{fig:TSP_Tensors}, the tensor $F(a)^i$ will be in charge of receiving whether the node $a$ has appeared 0 or 1 times between the qudits $0$ and $i-1$. The binary bond index has the semantics
\begin{equation}
    k=0 \iff \text{node } a \text{ has not appeared before},\qquad
    k=1 \iff \text{node } a \text{ has already appeared before}.
\end{equation}
With this convention, the following logics are fulfilled:
\begin{itemize}
    \item If node $a$ has appeared before: \begin{itemize}
        \item If qudit $i$ is in state $a$, it removes that combination.
        \item If qudit $i$ is not in state $a$, it does not remove that combination and communicates to the next layer tensor that node $a$ has appeared before.
    \end{itemize}
    \item If node $a$ has not appeared before: \begin{itemize}
        \item If qudit $i$ is in state $a$, it communicates to the next layer tensor that node $a$ has appeared before.
        \item If qudit $i$ is not in state $a$, it communicates to the next layer tensor that node $a$ has not appeared before.
    \end{itemize}
\end{itemize}
The final tensor is different: it also enforces that node $a$ has appeared exactly once by the end of the chain.

The tensors $F(a)^0$ (Fig.~\ref{fig:TSP_Tensors} a) and $F(a)^{\hat{N}-1}$ (Fig.~\ref{fig:TSP_Tensors} c) have indices $i,j$ of dimension $\hat{N}$ and index $k$ of dimension $2$, while the $F(a)^n$ (Fig.~\ref{fig:TSP_Tensors} b) in the rest of the cases, their indices $i,j$ still have dimension $\hat{N}$, while the indices $k,l$ have dimension $2$. These bond indices have the function of communicating whether a certain node $a$ of the TSP graph has appeared up to that point or not. Their nonzero elements are those in which $i=j$ and
\begin{equation}
    F(a)^0_{ijk} = \delta_{ij}\left(\mathbf{1}_{\{i=a,\;k=1\}}+\mathbf{1}_{\{i\neq a,\;k=0\}}\right)
\end{equation}
\begin{equation}
    F(a)^n_{ijkl} = \delta_{ij}\mathbf{1}_{\left\{l=k+\mathbf{1}_{\{i=a\}}\right\}}\mathbf{1}_{\{l\le 1\}}
\end{equation}
\begin{equation}
    F(a)^{\hat{N}-1}_{ijk} = \delta_{ij}\left(\mathbf{1}_{\{i=a,\;k=0\}}+\mathbf{1}_{\{i\neq a,\;k=1\}}\right)
\end{equation}
for pairs of different nodes. The intermediate tensor therefore implements the transition $l=k+\mathbf{1}_{\{i=a\}}$ while forbidding $l=2$, so the case $i=a$, $k=1$ is removed. The final tensor accepts only histories with final count exactly one, thereby excluding the incorrect case $i\neq a$, $k=0$ in which node $a$ never appears.

The number of nonzero entries is therefore
\begin{equation}
    \operatorname{nnz}\left(F^{(a),0}\right)=\hat{N},
    \qquad
    \operatorname{nnz}\left(F^{(a),n}\right)=2\hat{N}-1,
    \qquad
    \operatorname{nnz}\left(F^{(a),\hat{N}-1}\right)=\hat{N}.
\end{equation}
The dense intermediate filter tensor would have $4\hat{N}^2$ entries, while only $2\hat{N}-1$ are nonzero. If $i=a$, only the transition $0$ to $1$ is allowed; if $i\neq a$, the memory bit is copied.

\begin{lemma}\label{lem:exactly_once_filter}
Let $c_a(\vec{y})=|\{t:y_t=a\}|$. The contraction of the layer $F(a)$ on a configuration $\vec{y}$ is nonzero if and only if $c_a(\vec{y})=1$.
\end{lemma}

\begin{proof}
The initial tensor records whether $a$ appears in the first position. Each intermediate tensor propagates the memory bit that indicates whether $a$ has already appeared, and rejects any second appearance because the transition $k=1\to l=2$ is forbidden. Therefore, every surviving configuration has $c_a(\vec{y})\le 1$. The final tensor then removes the remaining case $c_a(\vec{y})=0$, because it accepts only $(i=a,k=0)$ or $(i\neq a,k=1)$. Hence the full layer is nonzero exactly when $c_a(\vec{y})=1$.
\end{proof}

Since the chain has $\hat{N}$ positions and takes values in an alphabet of size $\hat{N}$, it is enough to impose $c_a(\vec{y})=1$ for $\hat{N}-1$ distinct values. If $r$ is the unique unfiltered value, then
\begin{equation}
    c_r(\vec{y})=\hat{N}-\sum_{a\neq r}c_a(\vec{y})=\hat{N}-(\hat{N}-1)=1.
\end{equation}
Therefore, $\hat{N}-1$ layers already implement the same permutation constraint as
\begin{equation}
    R_{\vec{y},\vec{z}} =  |\epsilon_{y_0,y_1, \dots,y_{\hat{N}-1}}|^2\delta_{\vec{y},\vec{z}}.
\end{equation}

As we can observe, these tensors are highly sparse, since the only nonzero elements satisfy $i=j$; if $i=a$, only the transition $0$ to $1$ is allowed, whereas if $i\neq a$, the memory bit is copied. To give a numerical example, if $\hat{N}=10$, of the 400 elements ($10\times 10\times 2\times 2$) that any intermediate $F(a)^n$ would have, only 19 would be non-zero: one for $(a,0,1)$, 9 for the 9 possible values of b in $(b,0,0)$ and another 9 for the 9 possible values of b in $(b,1,1)$.

\subsection{Iterative resolution and contraction}\label{ssec:contraction}
We apply the iterative method of extraction, choosing randomly when encountering degenerate cases, and we obtain the different steps of the path. In each solving iteration, we solve the first step of a TSP without the nodes already obtained, with an initial condition at the last node obtained and the same final condition. In this way, in iteration $i$ we will create the tensor network for a problem with variables $\mathcal{N}=\hat{N}-i$.

\begin{proposition}\label{prop:iterative_correctness}
Assume that at some iteration the current prefix $p=(x_0,\dots,x_{r-1})$ can be extended to an optimal route. If the next value $x_r$ is chosen among the maximizers of $Z_p(a;\tau)$ in the limit $\tau\to\infty$, then the new prefix $(x_0,\dots,x_r)$ can also be extended to an optimal route. Consequently, repeating this rule until all positions are fixed produces an optimal route. In the presence of degeneracy, any maximizing branch in $I_p^\star$ is valid, so ties may be broken arbitrarily or at random.
\end{proposition}

\begin{proof}
The proof is by induction on the prefix length. For the empty prefix, there always exists at least one global optimum, so the claim holds. Assume it holds for a prefix $p$ that can be extended to an optimal route. By Proposition~\ref{prop:boltzmann_marginals}, every branch $a\notin I_p^\star$ is exponentially suppressed relative to every branch $b\in I_p^\star$ as $\tau\to\infty$. Hence any maximizer of $Z_p(a;\tau)$ in that limit belongs to $I_p^\star$, which means that the chosen value $x_r$ still admits an optimal continuation. The induction step follows, and after all positions are fixed the completed route is optimal.
\end{proof}

The preceding propositions are statements about the formal tensor-network expression and its zero-temperature limit. For a fixed prefix $p$ and candidate $a$, the relevant marginal free energy satisfies
\begin{equation}
    -\frac{1}{\tau}\log Z_p(a;\tau)
    \xrightarrow[\tau\to\infty]{}
    \min_{x\in \mathcal{F}_{p,a}} C(x).
\end{equation}
Thus, the mathematically exact extraction rule is the zero-temperature, exact-arithmetic, or tropical limit of the marginal comparison. The numerical implementation in Sec.~\ref{sec:experiments} instead uses finite $\tau$ and floating-point arithmetic, so it can fail when optimality gaps are small, degeneracies are present, or the numerical contrast between feasible branches is insufficient.

\begin{algorithmblock}[Sequential extraction by tensor marginals]\label{alg:marginal-extraction}
Input: edge or time-dependent costs, a fixed start/boundary node when required, the selected filter layers or constraints, and a value of $\tau$.
\begin{enumerate}
    \item Initialize an empty prefix  $p\leftarrow \emptyset$, the set of allowed nodes or states, and the corresponding filter memories.
    \item For each free position  $r$ in the route, build or update the tensor network conditioned on the current prefix  $p$.
    \item For every admissible candidate  $a$ at position  $r$, contract the tensor network with  $x_r=a$ fixed to obtain the marginal weight  $Z_p(a;\tau)$. This is performed for all values contracting the tensor network without fixing the variable index.
    \item Select  $x_r\in\arg\max_a Z_p(a;\tau)$, breaking ties arbitrarily or at random when the maximizers are degenerate.
    \item Append the selected value to the prefix, update the initialization support, and update or remove the filter layers whose counts have become fixed.
    \item After all free positions are selected, return the completed path or cycle, including the fixed boundary nodes when present.
\end{enumerate}
In exact arithmetic and in the limit $\tau\to\infty$, Proposition~\ref{prop:iterative_correctness} justifies this rule. At finite $\tau$ and finite precision, the same pseudocode describes the heuristic implementation evaluated below.
\end{algorithmblock}

The tensors of this new iteration tensor network $i$ will be the same as in the previous iteration $i-1$, but eliminate the possibility of choosing again a previously chosen node. To do this, we have two options. The first is to re-create all the tensors from scratch taking into account the new problem, which would imply a reduction of the dimensionality of each particular node but having to re-do all the creation and all the contractions. The second is to use the same tensors from the previous iteration, but making the `+' tensors have as nonzero values only those corresponding to the nodes not yet chosen. By doing this, the `$F$' filter layers of the nodes that have already appeared are unnecessary, since they will never be activated, so we can remove them to have a smaller and exponentially faster contracting network tensor. After that, we only have to change the second tensor of the minimization layer, $S^1$, which will now be the first one, so that it collects the information about the node chosen in the previous iteration, following the method presented in \cite{Tridiag}. Although the first method is more efficient in dimensionality of each of the individual tensors, the second method allows us to reuse intermediate calculations, so it is more advisable.

\begin{figure}[h]
  \centering
  \includegraphics[width=\linewidth]{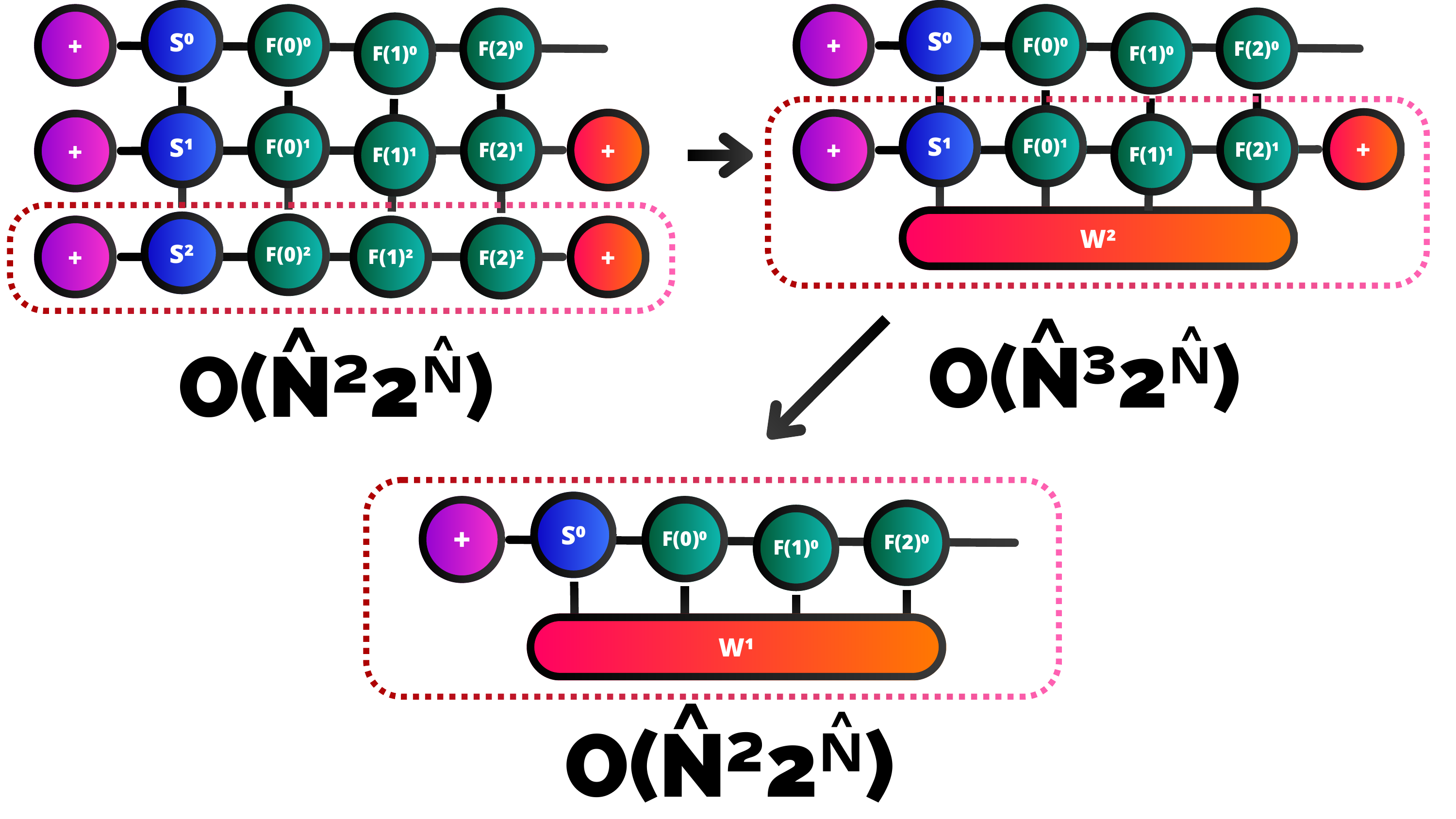}
  \caption{Contraction scheme for 3 variables with the cost of its steps for $\hat{N}$ variables.}
  \label{fig:Contraction}
\end{figure}

Our contraction scheme is presented in Fig.~\ref{fig:Contraction}. To reduce computational complexity and take advantage of intermediate calculations, we contract the tensor network by variables. First, we contract all the layers of the last variable, obtaining the tensor $W^{\hat{N}-1}$ with $\hat{N}$ indices and $\mathcal{O}\left(\hat{N}2^{\hat{N}}\right)$ elements. After that, we contract the $W^{\hat{N}-1}$ layer with the nodes of the previous variable, obtaining a new tensor $W^{\hat{N}-2}$ with the same number of indices and elements. We repeat this process with each variable until the entire tensor network is contracted. The elements of the tensor $W^i$ are $W^i_{s,k_0,k_1,\dots,k_{\mathcal{N}}}$, such that the first index $s$ is connected to the tensor $S^{i-1}$ above and the index $k_j$ is connected to the tensor $F(j)^{i-1}$.

\subsection{Computational complexity analysis}\label{ssec:complexity}

We will analyze the computational complexity of the contraction scheme. As the number of remaining free variables $\mathcal{N}$ in each iteration of the global extraction procedure decreases, we will calculate the complexity for a generic extraction iteration and then add it to obtain the complexity of the total number of iterations. We will assume that contracting a tensor of $N$ indices of dimension $n_i$ with another of $M$ indices of dimension $m_j$ through its first index has the usual computational cost of $\mathcal{O}\left(n_0\prod_{i,j=1,1}^{N-1,M-1} n_i m_j\right)$.

Contracting the last variable tensors to obtain $W^{\hat{N}-1}$ with $\mathcal{N}$ variables and a reduced problem with $\hat{N}=m$ free positions requires $\mathcal{O}\left(\hat{N}^2 2^{\mathcal{N}}\right)$ steps and contracting each tensor $W^i$ with the previous variable layers requires $\mathcal{O}\left(\hat{N}^3 2^{\mathcal{N}}\right)$ steps.

The cost of this first tensor $W^{\hat{N}-1}$ is given by the fact that in each step $i$ of the contraction we have a tensor $A^i$ with $i$ indices of dimension 2 connecting to the layers $F^{\hat{N}-2}$ of the previous variable, obtained by contracting the last $i$ tensors $F^{\hat{N}-1}$ and the `+' of the tracing. We will contract it with another $F$ tensor of 3 indices of dimension 2, $\hat{N}$ and $\hat{N}$, with a cost of $\mathcal{O}\left(\hat{N}^2 2^i\right)$. However, in the last step we will find that the tensor $S$ will already have been contracted with the initialization `+' tensor, having two physical indices of dimension $\hat{N}$. Therefore, the complexity of contracting it with $A^{\mathcal{N}-1}$ is $\mathcal{O}\left(\hat{N}^22^\mathcal{N}\right)$. Thus, the total complexity is $\mathcal{O}\left(\hat{N}^22^\mathcal{N}\right)$. The complexity of contracting it with the tensors of the next layer follows the same reasoning.

Contracting the entire tensor network requires contracting the tensors of the $\mathcal{N}$ variables, so the complexity is $\mathcal{O}\left(\hat{N}^3\mathcal{N} 2^\mathcal{N}\right)$ steps to determine one variable with a memory cost of $\mathcal{O}\left(\hat{N}^2 2^\mathcal{N}\right)$. Using the fact that each variable is less expensive than the previous one to compute, due to the iterative reduction of layers and nodes, we see that the complexity to solve an entire TSP of $\hat{N}$ variables is $\mathcal{O}\left(\sum_{\mathcal{N}=1}^{\hat{N}}\hat{N}^3\mathcal{N} 2^\mathcal{N}\right)$, resulting in a complexity of $\mathcal{O}\left(\hat{N}^4 2^{\hat{N}}\right)$ with a memory cost of $\mathcal{O}\left(\hat{N}^2 2^{\hat{N}}\right)$. This is the complexity of a dense layer-wise contraction, even though the tensors involved are highly sparse.

The reuse strategy reduces the cost of the subsequent extraction steps by reusing stored $W^i$ tensors. Under the contraction path analyzed here, the first full marginal contraction still costs $\mathcal{O}\left(\hat{N}^4 2^{\hat{N}}\right)$ and remains the dominant dense term, so reuse mainly reduces practical runtime and increases memory rather than changing the dense asymptotic time. During the contraction of the tensor network for the first variable, we can store each obtained tensor $W_i$. This requires increasing the memory cost to $\mathcal{O}\left(\hat{N}^3 2^{\hat{N}}\right)$. Thus, for the dense implementation considered here,
\begin{equation}
    T_{\mathrm{dense,reuse}}
    =
    \mathcal{O}\left(\hat{N}^4 2^{\hat{N}}\right).
\end{equation}

To determine the variable $i$, we take the tensor $W^{i+1}$ and create a new tensor $V^{i+1}$ which will be the tensor $W^{i+1}$ reduced with the solution already obtained. To construct it, each index $k_j$ of $W^{i+1}$ is fixed according to the filter memory induced by the already chosen prefix:
\begin{equation}
    k_j=
    \begin{cases}
        1, & j\in\{x_0,\dots,x_{i-1}\},\\
        0, & j\notin\{x_0,\dots,x_{i-1}\}.
    \end{cases}
\end{equation}
The physical states corresponding to already selected nodes are also removed from the initialization tensors, so those nodes cannot be selected again. The fixed values of $k_j$ only encode the filter memory induced by the already selected prefix. Performing this requires $\mathcal{O}\left(\hat{N}^2 2^\mathcal{N}\right)$ steps. After that, we contract the $V^{i+1}$ tensor with the layers of the variable $i$ corresponding to this step, removing the $F^i$ tensors of the nodes that have already appeared. Thus, the complexity of this variable becomes $\mathcal{O}\left(\hat{N}^2 2^\mathcal{N}\right)$ instead of $\mathcal{O}\left(\hat{N}^3 \mathcal{N}2^\mathcal{N}\right)$. Therefore, reuse does not change the dense worst-case order for the basic TSP under this analysis, although it makes the later extraction steps cheaper.

\paragraph{Sparse-local contraction of the same tensor network.}
The estimates above correspond to a dense layer-wise implementation of the contraction scheme in Fig.~\ref{fig:Contraction}. However, the elementary tensors used in the construction are highly sparse and mostly deterministic. The sparse-local implementation considered here does not change the tensor network or the contraction scheme: it uses the same local tensors $F$, $S$, and trace tensors, the same indices, and the same intermediate tensors $A^i$, $W^i$, and $V^i$ as in the dense implementation. It only exploits the zero pattern of the local tensors when performing the same contractions as in Fig.~\ref{fig:Contraction}. This should be distinguished from rewriting the method as a dynamic-programming recurrence: the tensors $A^i$, $W^i$, and $V^i$ have the same indices as before; only zero tensor entries are skipped. If the intermediate tensors $A^i$, $W^i$, and $V^i$ are still materialized densely, the memory bounds do not necessarily improve. Memory reductions can only be claimed when these intermediates are also stored sparsely, which requires a separate count of their produced nonzero support.

In the dense analysis, contracting an intermediate tensor with the next variable has dominant cost, written with the current iteration size $\mathcal{N}$, $\mathcal{O}\left(\hat{N}^3\mathcal{N}2^\mathcal{N}\right)$ for one marginal. This dense count treats the local update as if, for each current physical value, there were $\hat{N}$ independent next physical or signal choices to sum over. The actual local tensors do not have this freedom. For each physical value $i$, a filter $F^{(a)}$ permits only one memory update: $0$ to $1$ when $i=a$, and a copied memory bit when $i\neq a$. Similarly, the nearest-neighbor layer $S$ does not require an independent sum over signal indices; once the physical pair is fixed, the compatible signal propagation is fixed by the copy constraint. A sparse-local kernel therefore visits only these allowed local transitions, removing the extra local $\hat N$ scan from the dense kernel while leaving the global current-iteration filter-memory structure $2^\mathcal{N}$ unchanged. This sparse-local estimate is different from reuse: it reduces the arithmetic cost of the local contractions themselves, while reuse only avoids recomputing some later intermediates. Thus, for one marginal,
\begin{equation}
    T_{\mathrm{sparse-local}}^{\mathrm{one\ marginal}}(\mathcal{N})
    =
    \mathcal{O}\left(\hat{N}^2\mathcal{N}2^\mathcal{N}\right).
\end{equation}
Summing the iterative contractions gives
\begin{equation}
    T_{\mathrm{sparse-local}}^{\mathrm{TSP}}
    =
    \sum_{\mathcal{N}=1}^{\hat{N}}
    \mathcal{O}\left(\hat{N}^2\mathcal{N}2^\mathcal{N}\right)
    =
    \mathcal{O}\left(\hat{N}^32^{\hat{N}}\right).
\end{equation}
With reuse, the first sparse-local complete contraction still dominates, so
\begin{equation}
    T_{\mathrm{sparse-local,reuse}}^{\mathrm{TSP}}
    =
    \mathcal{O}\left(\hat{N}^32^{\hat{N}}\right).
\end{equation}
These bounds refer to arithmetic operations. If the intermediate tensors are still materialized densely, the memory bounds remain those of the dense analysis. Sparse storage of $W^i$, $A^i$, and $V^i$ may reduce memory in practice, but it requires a separate count of the nonzero support of the produced intermediates.

The binary filter-memory indices play a role analogous to subset states in dynamic programming. Thus, the goal of the construction is not to improve the worst-case exact complexity of Held-Karp for the classical TSP, but to provide a modular tensor-network representation in which constraints, variants, sparse-local kernels, and approximate layer removal can be handled within the same formalism.

We see that this method is asymptotically smaller than performing a blind brute-force search, which would require $\mathcal{O}\left(\hat{N}!\right)$ operations.

\subsection{Approximate approach}\label{ssec:approximation}
As we have seen in previous sections, the computational complexity of the different exact tensor-network contractions is exponential in the number of nodes. This formal scaling is preferable to a blind enumeration but should not be read as a general advantage over specialized exact dynamic-programming or branch-and-cut solvers; it may be excessive for sufficiently large industrial cases. The exponential computational cost is due to the constraint layers, as each gives an extra index to the intermediate $W^i$ tensors of the contraction. This makes it clear that if we reduce the number of layers, we can control the scaling of the heuristic implementation.

If during the execution of the finite-$\tau$ implementation, instead of applying all the constraint layers, we only apply a subset of them, we can obtain an approximate solution to the problem. During solving, we already partially impose the non-repetition constraint, making it so that at step $i+1$ none of the previously determined $i$ nodes can be activated. Given this, despite reducing the number of layers used, we can approximate a solution that satisfies the constraints. The main limitation of this method is that, in certain problems, there is a possibility that in the last steps of the solution only nonconnected nodes are available. However, if we choose correctly the layers to apply in each step of the extraction, this problem can be avoided.

To choose which rules are relevant to obtain the solution, we have two options. The first is to choose at each step a set of constraint layers at random from among the constraints that have not been imposed through initialization. The second is heuristically, by means of node closeness relations or based on previous failed solutions. Obviously, this version will not tend to obtain the global minimum, but it will tend to obtain sufficiently good solutions.

Another possibility to obtain approximate solutions is that, in the contraction scheme, instead of using the dense tensors to operate, we perform a compression in matrix product state (MPS) representation. Thus, both the intermediate $A^i$ tensors and the $W^i$ tensors are in approximate MPS representations, allowing us to require much less memory, approximating the system.

\FloatBarrier
\section{Generalized TSP cases}\label{sec:general_tsp}
\subsection{Different number of steps than number of nodes (DNSNN)}\label{ssec:dnsnn}
We can generalize the problem to other cases of industrial interest. The first is the case of having $N_s$ time steps, $N$ nodes, and each node $i$ can be visited between $N^0_i$ and $N^f_i$ times, inclusively:
\begin{equation}
    N_i^0\le c_i(\vec{x})\le N_i^f.
\end{equation}

In this case, we make a tensor network with the same structure as before, with $N_s$ qudits, but with a change in the filter layers. We generalize the tensors $F(a)$, which are used to make sure that each node has appeared exactly once in our solution, as tensors $F\left(a,N^0_a,N^f_a\right)$, which will make sure that node $a$ only appears between $N^0_a$ and $N^f_a$ times. They will follow the same logics as the $F$ tensors presented above, but with the difference that now they will only remove the combination if node $a$ has appeared $N^f_a$ times and its qudit is in $a$. Also, if it has appeared less than $N^0_a-1$ times until the last qudit, or if it has appeared $N^0_a-1$ and the last qudit is not in $a$. In addition, they will communicate to the next tensor how many times $a$ has appeared up to that qudit.

Tensors $F\left(a,N^0_a,N^f_a\right)^0$ (Fig.~\ref{fig:TSP_Tensors} d) and $F\left(a,N^0_a,N^f_a\right)^{N_s-1}$ (Fig.~\ref{fig:TSP_Tensors} f) have indices $i,j$ of dimension $N$ and index $k$ of dimension $N^f_a+1$, while $F\left(a,N^0_a,N^f_a\right)^n$ (Fig.~\ref{fig:TSP_Tensors} e) in the rest of the cases, their indices $i,j$ have dimension $N$, while the indices $k,l$ have dimension $N^f_a+1$. These bond indices have the function of communicating how many times a certain node $a$ of the TSP graph has appeared up to that point. The bond indices can have a smaller dimension, adapting to the step in which we are, but not if we want to reuse intermediate steps. However, we will not consider this adaptation in the paper, as it would only complicate the analysis. Their nonzero elements are those in which $i=j$ and
\begin{equation}
    F(a,N^0_a,N^f_a)^0_{ijk}=\delta_{ij}\left(\mathbf{1}_{\{i=a,\;k=1\}}+\mathbf{1}_{\{i\neq a,\;k=0\}}\right)
\end{equation}
\begin{equation}
    F(a,N^0_a,N^f_a)^n_{ijkl}=\delta_{ij}\mathbf{1}_{\left\{l=k+\mathbf{1}_{\{i=a\}}\right\}}\mathbf{1}_{\{l\le N^f_a\}}
\end{equation}
\begin{equation}
    F(a,N^0_a,N^f_a)^{N_s-1}_{ijk}=\delta_{ij}\left[\mathbf{1}_{\{i=a\}}\mathbf{1}_{\{N_a^0\le k+1\le N_a^f\}}+\mathbf{1}_{\{i\neq a\}}\mathbf{1}_{\{N_a^0\le k\le N_a^f\}}\right]
\end{equation}

Again, the tensors are very sparse. The basic TSP is recovered by taking $N_a^0=N_a^f=1$ for the filtered nodes, in which case the final DNSNN tensor reduces exactly to the corrected ``appears exactly once'' filter of Sec.~\ref{ssec:construction}.

As before, we use the iterative method to obtain the different nodes in different time steps. Here, after determining in one step $i$ the node $a$, in the next iteration, we reduce $N^0_a$ and $N^f_a$ by one unit. If $N^0_a$ reaches 0, it stops decreasing. If $N^f_a$ reaches 0, we remove the filter layer from that node and eliminate that possibility in the `+' tensors. At an iteration with $r$ positions already fixed and remaining horizon $h=N_s-r$, if $N^0_a=0$ and $N^f_a\ge h$, the filter for node $a$ no longer imposes a restriction on the still-free positions and can be omitted. This condition depends on the actual number of remaining stops of the variant, not automatically on $N-1$. The contraction scheme is the same as the one described for the original TSP.

To analyze the computational complexity, without loss of generality, let us assume that each node can appear up to a uniform number of $N^f$ times, and define the counting-index dimension
\begin{equation}
    D=N^f+1.
\end{equation}
Following the reasoning of the previous algorithm, the complexity of contracting the last tensor $W$ is $\mathcal{O}\left(N^2D^N\right)$, contracting it with the previous variable layers has a complexity of $\mathcal{O}\left(N^3D^N\right)$, and the complexity of contracting the entire tensor network would be $\mathcal{O}\left(N^3N_sD^N\right)$. To determine the variables $N_s$, we would have a complexity of $\mathcal{O}\left(N^3N_s^2D^N\right)$ with a memory cost of $\mathcal{O}\left(N^2D^N\right)$ in the worst case.

In this case, we can also use the reuse of intermediate calculations, this time the tensors $W^{i+1}$ have indices $k_j$ of dimension $D$. For this process, instead of creating a tensor $V^{i+1}$, we redefine the tensors $F(j)^i$, so that if node $j$ has appeared $n_j$ times in the solution, the first $n_j$ elements of the bond index are null. This is to tell the tensor $W^{i+1}$ that no such node can have appeared less than $n_j$ times. Thus, the computational complexity of these steps ranges from $\mathcal{O}\left(N^3N_sD^N\right)$ to $\mathcal{O}\left(N^2D^N\right)$. Storing all reusable tensors raises the memory cost to $\mathcal{O}\left(N^2N_sD^N\right)$, while the total computational complexity of determining all the variables is reduced to $\mathcal{O}\left(N^3N_sD^N\right)$.

\subsection{Non-markovian TSP (NMTSP)}
The second case we will study is the TSP with memory. This problem consists of a generalization in which the cost of a step depends not only on the previous step, but also on the previous ones up to a certain term. In this case, we stop having a QUDO case and we move to a HODO (Higher Order Discrete optimization) case.

For a number of steps $K$ to be considered in each time interval, we have the cost function.
\begin{equation}
    C(\vec{x}) = \sum_{t=0}^{N-1} E_{t, x_{t+1}, x_t, x_{t-1}, \dots, x_{t-K+1}},
\end{equation}
being $E_{t,i,j_0, \dots, j_{K-1}}$ the cost of moving from node $j_0$ to node $i$ at step $t$ with $j_{1}, \dots, j_{K-1}$ being the last $K$ nodes. If $t<K$, we only use the nodes until $x_0$ or $x_{N-n}=x_{-n}$ if we want the return condition.

\begin{figure}[h]
  \centering
  \includegraphics[width=\linewidth]{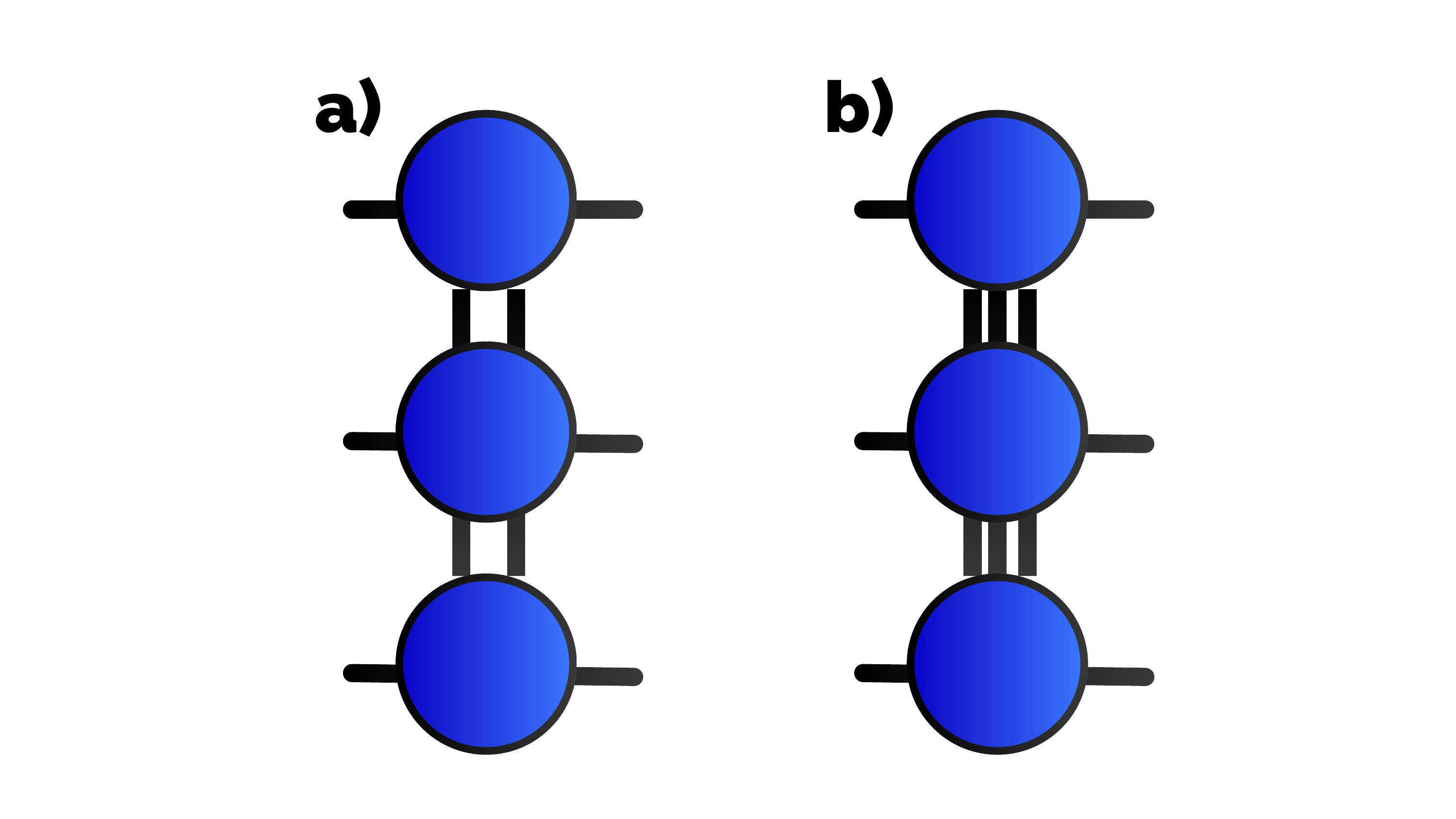}
  \caption{MPO layer for 3 multiple bond index tensors.}
  \label{fig:TSP_Multibond}
\end{figure}

For this problem, we change the tensor layer $S$ to a tensor MPO layer $S(K)$ where for each bond index $k,l$ we use $K$ bond indices $k_m, l_m$, as in Fig.~\ref{fig:TSP_Tensors}. Each of them passes the signal from a previous node and an imaginary time evolution is applied according to the signal obtained through the input indices. Thus, the non-zero elements of these tensors are those with $i=j=l_0$ and $l_{m+1}=k_m \ \forall m\in[0,K-2]$. Equivalently, at site $t$, with incoming memory $\mathbf{k}=(k_0,\dots,k_{K-1})$ and outgoing memory $\mathbf{l}=(l_0,\dots,l_{K-1})$, the layer can be written as
\begin{equation}
    S^{(K),t}_{i,j,\mathbf{k},\mathbf{l}}
    =
    \delta_{ij}\,
    \mathbf{1}_{\{l_0=i\}}
    \prod_{m=0}^{K-2}\mathbf{1}_{\{l_{m+1}=k_m\}}
    e^{-\tau E_{t,i,k_0,\dots,k_{K-1}}},
\end{equation}
with the unavailable boundary-memory entries omitted or filled according to the chosen open-route or return convention. They will be connected as in Fig.~\ref{fig:TSP_Multibond}.

The value of its elements is analogous to that described in general terms in \cite{Tridiag}. The rest of the method is analogous to that used in the previous cases and can be combined with the generalization of Sec.~\ref{ssec:dnsnn}, because the former only changes the filtering layers, while the present variant only changes the evolution layer.

Its computational complexity without reuse is $\mathcal{O}\left(N^{2K+2}2^N\right)$ with memory cost $\mathcal{O}\left(N^{K+1}2^N\right)$ in the simple case and $\mathcal{O}\left(N^{2K+1}N_s^2D^N\right)$ with memory cost $\mathcal{O}\left(N^{K+1}D^N\right)$ in the version of Sec.~\ref{ssec:dnsnn}.

For reuse we do the same as indicated for the original problem, obtaining computational complexity $\mathcal{O}\left(N^{2K+2}2^N\right)$ with memory cost $\mathcal{O}\left(N^{K+2}2^N\right)$ in the simple case and $\mathcal{O}\left(N^{2K+1}N_sD^N\right)$ with memory cost
$\mathcal{O}\left(N^{K+1}D^N\max\left(N,N_s\right)\right)$ in the version of Sec.~\ref{ssec:dnsnn}.

\subsection{Bottleneck TSP (BTSP)}
The variant we are going to solve now is the TSP with bottleneck. The problem is to find the route that minimizes the cost of the highest-cost edge of the route. The applications of this problem are multiple, such as planning bus routes or designing plate drilling routes. 

The formulation of this problem is as follows.
\begin{gather}
    \vec{x}_{opt} = \arg\min_{\vec{x}}\left(  \max_{t=0}^{N-1} E_{x_t,x_{t+1}}\right)\\
    x_t\in[0, N-1]\ \forall t\in [0, N-1]\nonumber\\
    x_t \neq x_{t'} \qquad \forall t\neq t' \in [0, N-1]\nonumber,
\end{gather}
To solve this problem, we use the same scheme as in Fig.~\ref{fig:TensorNetworkTSP}, but changing the $S$ tensor layer to a $Z$ tensor layer. Now, this layer will be in charge of communicating which is the highest displacement cost. The final tensor will be responsible for applying the imaginary time evolution. For the same reason, we need that instead of having a single index bond that communicates the node of the previous step, we have two that communicate the previous node and the maximum cost that has appeared until that step.

For this method, we assume that all elements of the tensor $C$ are positive integers. Let
\begin{equation}
    Q=\{0\}\cup\{E_{a,b}:a,b\in V\},
\end{equation}
where $0$ is the initial state of the bottleneck register, and define
\begin{equation}
    M_c:=|Q|.
\end{equation}
If the costs take values in $\{1,\dots,M_{\max}\}$, then $Q\subseteq\{0,1,\dots,M_{\max}\}$ and, in the dense integer case, $M_c=M_{\max}+1$. The generalization to rational numbers and the approximation to integers with truncation are straightforward. For the sake of simplicity, we also assume the returning condition; generalization to other cases is straightforward.

Its physical indices $i,j$ are of dimension $\hat{N}$, its bond indices $k,l$ for the node signal in the previous step are of dimension $\hat{N}$, and its bond indices $q,p$ for carrying the current bottleneck value take values in $Q$, so their dimension is $M_c$.
The non-zero elements of these tensors are those in which $i=j=k$ for $Z^0$, $i=j=l$ for the $Z^n$ and $i=j$ for the $Z^{N-2}$. These nonzero elements are
\begin{equation}
    Z^0_{ijkq}=\delta_{ij}\delta_{ki}\mathbf{1}_{\{q=E_{x_{\mathrm{start}},i}\}}
\end{equation}
\begin{equation}
    Z^n_{ijklqp}=\delta_{ij}\delta_{li}\mathbf{1}_{\{p=\max(q,E_{k,i})\}}
\end{equation}
This transition is deterministic: once $(q,k,i)$ is fixed, there is only one compatible outgoing bottleneck value $p$.
\begin{equation}
    Z^{N-2}_{ijkq}= \delta_{ij}e^{-\tau \max(q, E_{k,i}, E_{i,x_{\mathrm{start}}})}
\end{equation}
where $x_{\mathrm{start}}$ is the fixed initial/final node and $\tau$ is the decay hyperparameter.

Another variation would be to find the path that maximizes the cost of the lowest-cost edge of the path. This variant is useful for scheduling metalworking steps in aircraft manufacture, where we need to avoid accumulating heat in steps close together in space and time.

The rest of the extraction procedure is exactly the same as the previous ones, but changing the $Z^0$ so that its nonzero elements are at $q=\max(E_{x_{m-1},x_m},E^{\max}_m)$, where $m$ is the position currently being extracted and $E^{\max}_m$ is the bottleneck value accumulated by the already fixed prefix.

To solve this version, we only have to do the same as in the minimization version, but changing all $\max$ by $\min$ and vice versa and using a $\tau<0$ to maximize.

Its computational complexity is $\mathcal{O}\left(\hat{N}M_c 2^{\hat{N}}\max\left(\hat{N}, M_c \right)\right)$, with a cost in memory $\mathcal{O}\left(\hat{N}M_c 2^{\hat{N}}\right)$. With reuse, the computational complexity is $\mathcal{O}\left(\hat{N}M_c 2^{\hat{N}}\max\left(\hat{N}, M_c \right)\right)$, with a cost in memory $\mathcal{O}\left(\hat{N}^2M_c 2^{\hat{N}}\right)$.

\subsection{\texorpdfstring{Group-constrained TSP (Politician TSP)}{Group-constrained TSP (Politician TSP)}}
Another variant with industrial utility is the group-constrained TSP, also known here as the traveling politician problem, which deals with states that have several cities each and you have to visit exactly one city in each state. This can be abstracted as a graph with $N$ nodes of different $M$ classes, so that we have to visit one node of each class. Its industrial utility includes the more efficient cutting of types of sheets of paper from a larger sheet.

The formulation of this problem is as follows.
\begin{gather}
    \vec{x}_{opt}=\arg\min_{\vec{x}}C(\vec{x}),\nonumber\\
    V=\{0,\dots,N-1\},\qquad g:V\to\{0,\dots,M-1\},\nonumber\\
    \vec{x}=(x_0,\dots,x_{M-1}),\qquad x_t\in V\ \forall t\in [0,M-1],\nonumber\\
    \{g(x_0),\dots,g(x_{M-1})\}=\{0,\dots,M-1\},\nonumber\\
    g(x_t)\neq g(x_{t'})\qquad \forall t\neq t' \in [0,M-1],\nonumber\\
    C(\vec{x}) = \sum_{t=0}^{M-1} E_{t,x_t, x_{t+1}},\qquad x_M=x_0,
\end{gather}
for the closed-route version. For an open path, the cost is
\begin{equation}
    C(\vec{x}) = \sum_{t=0}^{M-2} E_{t,x_t, x_{t+1}}.
\end{equation}
If each city belongs to a unique class, then forbidding repeated classes already prevents repeated cities, so no additional non-repetition constraint is needed.

To solve this problem, we change the filter layers $F$ in a very simple way. Instead of associating each layer to a node, we associate it to a class. The tensor chain now has length $M$, while each physical index still has dimension $N$ because every position can choose any of the $N$ cities. As in the basic TSP, only $M-1$ class filters are required, because if $M-1$ classes appear exactly once, the remaining class is forced by counting. Each class filter activates signal 1 and filters for all the states belonging to that class.

The rest of the algorithm is exactly the same, except that each time a node of a class appears, we remove from later positions the initialization support and the class filter not only of that node but of all nodes in the same class.

Its complexity is $\mathcal{O}\left(N^3 2^M\right)$ with a memory cost of $\mathcal{O}\left(N^2 2^M\right)$ without reuse and $\mathcal{O}\left(N^3 2^M\right)$ with a memory cost of $\mathcal{O}\left(N^3 2^M\right)$ with reuse. Here, the polynomial dependence on $N$ comes from the physical dimension of the sites, whereas the exponential factor $2^M$ comes from the $M-1$ class filters and the fact that the chain has $M$ steps.

\subsection{TSP with precedence (TSPP)}
Another interesting case is the TSP in which we have a precedence rule between nodes. That is, certain nodes have to appear earlier in the route than others. This additional rule is straightforward to implement in the first version of the algorithm by slightly modifying the tensors $F(a)$. We will see how to do it with an example.

A minimal general formulation is obtained by defining a precedence set
\begin{equation}
    P\subseteq V\times V,
\end{equation}
where $(u,v)\in P$ means that node $u$ must appear before node $v$. A feasible route must satisfy
\begin{equation}
    \operatorname{pos}(u)<\operatorname{pos}(v)
    \qquad
    \forall (u,v)\in P.
\end{equation}
Tensorially, each node $u$ already has a memory filter indicating whether it has appeared. A node $v$ is allowed at a given position only if all nodes $u$ with $(u,v)\in P$ have already appeared. This can be implemented by adding local projectors coupled to the memory bits of the corresponding predecessor filters. If at least one required predecessor memory bit is zero, the physical value $v$ is suppressed at that position. When several predecessors point to the same node $v$, all of their memory bits must be active.

Writing $\mathrm{Pred}(v)=\{u:(u,v)\in P\}$ and denoting by $k_u$ the memory bit that indicates whether predecessor $u$ has already appeared, an explicit local projector is
\begin{equation}
    P_v(i,\{k_u\}_{u\in \mathrm{Pred}(v)})
    =
    \mathbf{1}_{\{i\neq v\}}
    +
    \mathbf{1}_{\{i=v\}}
    \prod_{u\in \mathrm{Pred}(v)} k_u.
\end{equation}
Thus, if the physical value is $v$, all predecessor memory bits must already be active; otherwise the local amplitude is set to zero.

Let us imagine that we have the rule that node 4 has to appear before node 7 in the path. To take this into account, the local projector coupled to the memory bit of $F(4)$ suppresses the physical value $i=7$ whenever that memory bit is still $0$. Equivalently, the precedence projector itself forbids $i=7$ at the first position because the memory bit of node $4$ is initially inactive; it also forbids completions in which node $4$ would appear only after node $7$. These restrictions come from the precedence projector and its boundary memory state, not from an unexplained initialization rule. The rest of the construction is exactly the same.

\subsection{\texorpdfstring{ONCE Job Reassignment proof of concept}{ONCE Job Reassignment proof of concept}}\label{sec:once}
We now present a real use case in which we have applied this method \cite{ONCE}. This is the case of the Job Reassignment Problem (JRP) developed for the Organizaci\'on Nacional de Ciegos Espa\~noles (ONCE).

The job re-assignment problem consists of, given a set of workers assigned each to one of a set of jobs, and a set of vacant jobs, finding whether any of the assigned workers should move to the vacant jobs. Each job has its own quality score, given by its profitability, and an affinity with each of the workers, given by the suitability of this worker occupying that job. The objective is to obtain an assignment that increases the sum of productivity and affinity in the group of workers.

Since the whole process is extensively analyzed in the original paper, we focus on the subproblem, where given a set of workers and vacant positions, we have to choose which position to send them (if any). The constraint is that no two workers can be sent to the same job. Each worker can only be assigned to one job, but that will be ensured by the encoding of the variables and will not be a restriction we need to impose to our tensor network.

The solution is encoded as a vector $\vec{x}$ of integer values, where $x_i$ is the vacant position to which the worker $i$ is sent. Moreover, $x_i=0$ means that the worker $i$ stays in his position without moving to any of the vacant ones. Its cost function is
\begin{equation}
    C(\vec{x})=\sum_{i=0}^{N-1} \left[ c^P(P^C_i-P^V_{x_i})+c^A(A^C_{i,i}-A^V_{x_i,i})\right],
\end{equation}
being $P^V_{x_i}$ the quality of the $x_i$ vacant job, $P^C_i$ the quality of the current $i$ worker job, $A^V_{x_i,i}$ the affinity of the worker $i$ in the vacant job $x_i$, $A^C_{i,i}$ the affinity of the worker $i$ in its current job, and $c^P$ and $c^A$ the relative weights of productivity and affinity.

The value $x_i=0$ is a repeatable dummy ``no move'' state, not a physical vacancy. Equivalently, indexing real vacancies by $v\ge 1$, the local costs can be written as
\begin{equation}
    c_{i,0}=0,\qquad
    c_{i,v}=c^P(P^C_i-P^V_v)+c^A(A^C_{i,i}-A^V_{v,i}),\quad v\ge 1.
\end{equation}
This notation separates the no-move option from the ordinary vacancy states and avoids applying vacancy-specific quantities to the dummy value.

Therefore, we can see that our problem can be expressed as a general TSP problem as in Eq.~\eqref{eq:gen_TSP_QUDO} in which we have only the $E^0$ term. 

In the subproblem considered here, the cost is linear and the only coupling constraint is that a nonzero vacancy cannot be assigned to more than one worker. The state $x_i=0$, meaning that worker $i$ stays in the current position, may be repeated arbitrarily many times. Therefore, this subproblem can also be seen as a rectangular assignment problem, or equivalently as a bipartite matching problem with a repeated dummy option representing the decision of not moving a worker. Specialized classical solvers for assignment or min-cost-flow problems are polynomial for this structure. Hence, the JRP experiment should be interpreted as a validation of the tensor-network construction and of its modular industrial integration with constraints, rather than as a claim of computational advantage over assignment-specific algorithms.

Because this subproblem has only the linear term, the $S$ tensor layer is not needed; the imaginary-time factor is placed in the local `+' tensor of each variable. In addition, since several workers may not move jobs, we also remove the filter layer of $x_i=0$. The other filter layers will have $N^0=0$ and $N^f=1$, so that each nonzero vacancy may be chosen at most once. After each choice step of $x_i$, if we obtain $x_i\neq 0$, we remove that state from the `+' tensors and its corresponding filter layer. If, at any step, we run out of free vacancies, we stop the extraction and fill the remaining components of $\vec{x}$ with 0.

The computational complexity to solve a subproblem for $I$ vacant jobs and $J$ workers is $\mathcal{O}\left(J^2I^22^{I+1}\right)$ with a memory cost of $\mathcal{O}\left(I\  2^{I+1}\right)$. With reuse of intermediate calculations, we can achieve a complexity of $\mathcal{O}\left(JI^22^{I+1}\right)$ with a memory cost of $\mathcal{O}\left(JI 2^{I+1}\right)$.

Due to the symmetry of the problem, since assigning workers to jobs and jobs to workers is equivalent, in the case of $I>J$, we only have to run the same algorithm with the cost tensor $C_{i,x_i}$ changed so that now the elements $x_i$ are the worker we send to the vacant $i$ and $x_i=0$ means that no worker is sent to this vacant. Thus, the computational complexity is $\mathcal{O}\left(I^2J^22^{J+1}\right)$ with a memory cost of $\mathcal{O}\left(J\  2^{J+1}\right)$. With reuse of intermediate calculations, we can achieve a complexity of $\mathcal{O}\left(J^2I2^{J+1}\right)$ with a memory cost of $\mathcal{O}\left(JI  2^{J+1}\right)$.

The same JRP tensor network also benefits from sparse-local contractions, since the vacancy filters have deterministic counting transitions and the state $x_i=0$ has no filter layer. For $I\le J$, this removes one factor $I$ from the dense local tensor contractions while preserving the same tensor network and the same contraction trajectory:
\begin{gather}
    T_{\mathrm{sparse-local,no\ reuse}}^{\mathrm{JRP}}
    =
    \mathcal{O}\left(J^2I2^I\right),
    \qquad
    T_{\mathrm{sparse-local,reuse}}^{\mathrm{JRP}}
    =
    \mathcal{O}\left(JI2^I\right).
\end{gather}
For $I>J$, the same worker/vacancy symmetry described above applies. The assignment interpretation remains useful as a classical baseline for the linear JRP subproblem, but it is separate from the tensor-network sparse-local contraction analysis.

\subsection{Sparse-local contraction of the tensor networks}
The generalized tensor networks also benefit from sparse-local kernels, but the interpretation remains the same as in Sec.~\ref{ssec:complexity}: we do not introduce new state variables or a new dynamic-programming representation. The tensor network, indices, and contraction trajectory are unchanged; only locally zero entries are skipped. Unless stated otherwise, the memory bounds remain those of the dense analysis when $A^i$, $W^i$, and $V^i$ are still materialized densely.

For DNSNN, let
\begin{equation}
    D=N^f+1.
\end{equation}
The intermediate counting filter satisfies
\begin{equation}
    l=k+\mathbf{1}_{\{i=a\}},
    \qquad
    l\le N^f.
\end{equation}
Therefore, for an intermediate filter tensor,
\begin{equation}
    \operatorname{nnz}\left(F^{(a,N_a^0,N_a^f),n}\right)
    =
    (N-1)D+(D-1)
    =
    ND-1.
\end{equation}
This is $\mathcal{O}(ND)$, instead of the dense size $\mathcal{O}(N^2D^2)$. Along the same contraction trajectory, the sparse-local kernels reduce the factor $N^3$ in the dense DNSNN time estimate to $N^2$, while the factor $D^N$ from the counting-filter indices remains:
\begin{gather}
    T_{\mathrm{sparse-local,no\ reuse}}^{\mathrm{DNSNN}}
    =
    \mathcal{O}\left(N^2N_s^2D^N\right),\\
    T_{\mathrm{sparse-local,reuse}}^{\mathrm{DNSNN}}
    =
    \mathcal{O}\left(N^2N_sD^N\right).
\end{gather}
With dense intermediate tensors, the memory bounds remain those of the dense analysis.

For NMTSP, the same sparse-local principle applies to the $K$-memory layer $S^{(K)}$, because the memory-shift constraints are deterministic. A detailed count of the resulting polynomial prefactors depends on the precise implementation of $S^{(K)}$ and is left out of the summary table.

For BTSP, the bottleneck tensor $Z$ transports the accumulated maximum through the deterministic transition
\begin{equation}
    p=\max(q,E_{k,i}).
\end{equation}
Thus, for each $(q,k,i)$ there is at most one allowed value of $p$, and the sparse-local kernel does not scan the dense pair $(q,p)$. Writing $M_c=|Q|$,
\begin{equation}
    T_{\mathrm{sparse-local}}^{\mathrm{BTSP}}
    =
    \mathcal{O}\left(\hat{N}^2M_c2^{\hat{N}}\right).
\end{equation}
The dense memory bounds are retained unless the intermediate tensors are also stored sparsely.

For PTSP, the chain length is $M$, the physical dimension is $N$, and the class filters are binary and sparse. The dense analysis gives $\mathcal{O}(N^32^M)$. With sparse-local contractions, copy constraints and class activations do not scan dense physical combinations, giving
\begin{equation}
    T_{\mathrm{sparse-local}}^{\mathrm{PTSP}}
    =
    \mathcal{O}\left(N^22^M\right).
\end{equation}
The factor $2^M$ still comes from the class-filter indices. The memory bounds remain the dense ones unless sparse storage of the produced intermediates is analyzed.

For TSPP, precedence filters are also sparse-local projectors, so the same sparse-local implementation reduces polynomial prefactors in the contraction. A sharper complexity depending on the precedence structure is left for future work.

\FloatBarrier
\section{Experiments}\label{sec:experiments}

The experimental study is organized in three reproducible benchmark stages intended as small-instance diagnostics rather than as a definitive solver competition. First, we calibrate the imaginary-time parameter $\tau$ of the TN solver using a calibration split only, and then evaluate the effect of the selected schedule on a disjoint evaluation split. The main results below use the final wide calibration with an integer sweep over $\tau=\{1,\dots,40\}$ followed by local refinement. Second, with this calibrated schedule fixed, we compare TN full-filter mode with representative classical references: Held-Karp, OR-Tools~\cite{ortools}, and NetworkX greedy and simulated annealing heuristics~\cite{networkx}. Third, we perform a restriction-layer ablation to isolate the empirical effect of retaining only a fraction of the layers that enforce the TSP non-repetition constraints. This order separates parameter selection, external comparison, and internal mechanism analysis while keeping the calibration and evaluation data disjoint.

\subsection{\texorpdfstring{Experimental setup}{Experimental setup}}\label{ssec:experimental-setup}

All three stages use a deliberately controlled small-instance benchmark suite, where exact optima can be computed for every instance. This makes it possible to report objective quality metrics for the tours returned by the TN solver. The suite contains three synthetic TSP families: Euclidean symmetric TSP instances, random complete symmetric TSP instances, and random complete asymmetric TSP instances. The calibration split uses seeds $\{0,1,2,3,4\}$, whereas the evaluation split uses seeds $\{5,6,7,8,9\}$; these two partitions are disjoint.

For each tested size and split, each of the three instance families contributes five instances. The calibration and solver-comparison stages use $n\in\{5,6,7,8,9,10,12\}$. To quantify the effect of calibration, we compare TN full-filter mode with $\tau=1$ against TN full-filter mode with the calibrated schedules on the small-instance subset with known optima $n\in\{5,6,7,8,9,10\}$. This gives 90 evaluation instances for the small-instance calibration check and 105 evaluation instances for the classical comparison. The layer-ablation stage is run separately for $n\in\{8,10,12\}$, layer ratios $\rho\in\{0,0.1,0.25,0.5,0.75,1\}$, and five repeats per setting.

All solution-quality metrics are computed relative to the exact Held-Karp optimum $C^\star$. Held-Karp is therefore used as the exact baseline for objective values, not as a tuned runtime competitor. For a returned tour of cost $C$, the reported relative gap is $(C-C^\star)/C^\star$. A solution is counted as optimal only if it is a valid tour and its absolute relative gap is at most $10^{-9}$. We summarize each group by the optimal-solution rate, valid-route rate, finite-edge rate, median gap, mean gap, 95th-percentile gap, and worst gap. Unless explicitly varied in the ablation, TN full-filter mode means that all restriction layers are used. Runtime values are shown only as implementation diagnostics: they include Python overhead and solver-wrapper overhead, and they are not used here to claim computational superiority over specialized classical solvers.

\subsection{\texorpdfstring{Tau calibration and effect of calibration}{Tau calibration}}\label{ssec:tau-calibration}

The imaginary-time parameter $\tau$ controls the contrast between low-cost and high-cost tours in the Boltzmann weights used by the TN solver. A value that is too small leaves competing tours insufficiently separated, whereas a value that is too large may amplify numerical effects. We therefore calibrate $\tau$ by problem size, using one unified value for all three instance families at each $n$. Calibration is performed in two stages. The first stage sweeps integer values $\tau\in\{1,\dots,40\}$. The second stage refines around the first-stage winner with step size $0.2$, using ten values below and ten values above it. The selection criterion first maximizes the optimal-solution rate, then breaks ties by smaller median gap, smaller mean gap, and finally by the smaller stable $\tau$.

\begin{figure}[t]
    \centering
    \includegraphics[width=0.7\linewidth]{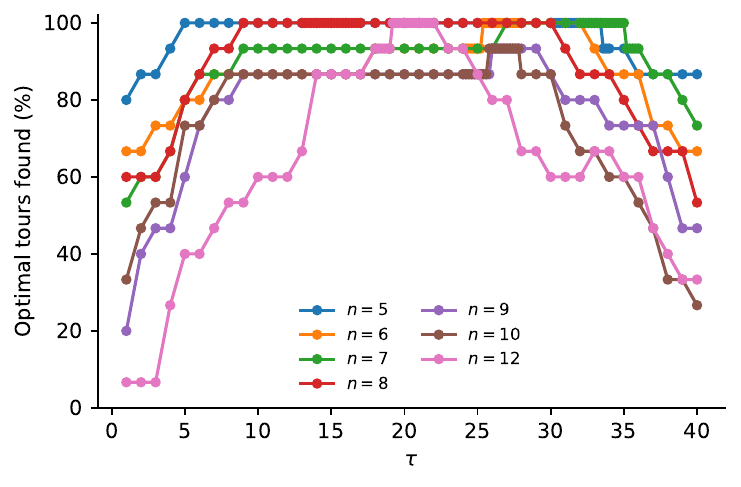}
    \caption{Optimal-solution rate of the TN solver during the wide $\tau=\{1,\dots,40\}$ calibration, with one curve for each tested problem size.}
    \label{fig:tau-optimal-rate}
  \end{figure}

  \begin{table}[t]
    \centering
    \caption{Selected $\tau$ value and calibration-split summary metrics for each tested problem size under the final $\tau=\{1,\dots,40\}$ calibration.}
    \footnotesize
    \begin{tabular}{rrrr}
        \toprule
        $n$ & selected $\tau$ & optimal rate & mean gap \\
        \midrule
        5 & 31.4 & 100.00\% & 0.000\% \\
        6 & 25.4 & 100.00\% & 0.000\% \\
        7 & 34.0 & 100.00\% & 0.000\% \\
        8 & 14.6 & 100.00\% & 0.000\% \\
        9 & 26.0 & 93.33\% & 0.006\% \\
        10 & 25.8 & 93.33\% & 0.045\% \\
        12 & 19.2 & 100.00\% & 0.000\% \\
        \bottomrule
    \end{tabular}
    \label{tab:tau-calibration}
\end{table}

\begin{figure}[t]
  \centering
  \includegraphics[width=0.5\linewidth]{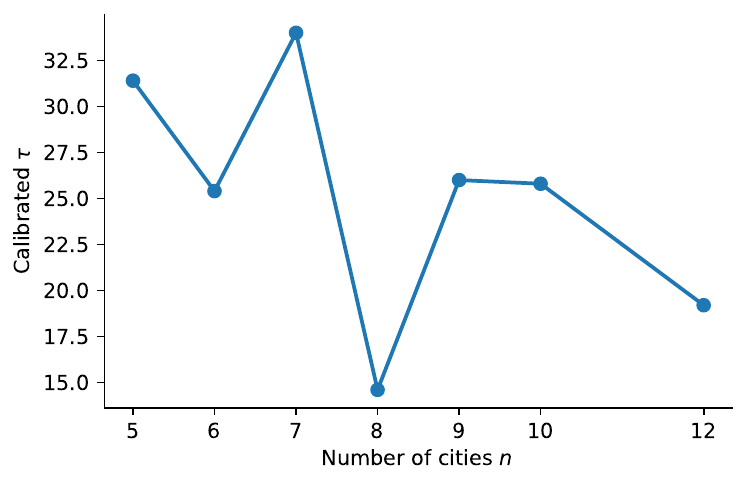}
  \caption{Selected calibrated $\tau$ schedule as a function of the TSP size.}
  \label{fig:tau-schedule}
\end{figure}

The selected values are $\tau=\{31.4,25.4,34.0,14.6,26.0,25.8,19.2\}$ for $n=\{5,6,7,8,9,10,12\}$, respectively.
The sweep in Fig.~\ref{fig:tau-optimal-rate} shows why we use this explicit lookup table. The wider scan shows that accuracy improves as $\tau$ increases into a favorable numerical region, but good values of $\tau$ remain size-dependent and the response is not monotone in $n$. The schedule in Table~\ref{tab:tau-calibration} and Fig.~\ref{fig:tau-schedule} should therefore be read as an empirical calibration schedule for the tested sizes, not as evidence for a reliable $\tau$ scaling law.

\begin{figure}[t]
    \centering
    \includegraphics[width=\linewidth]{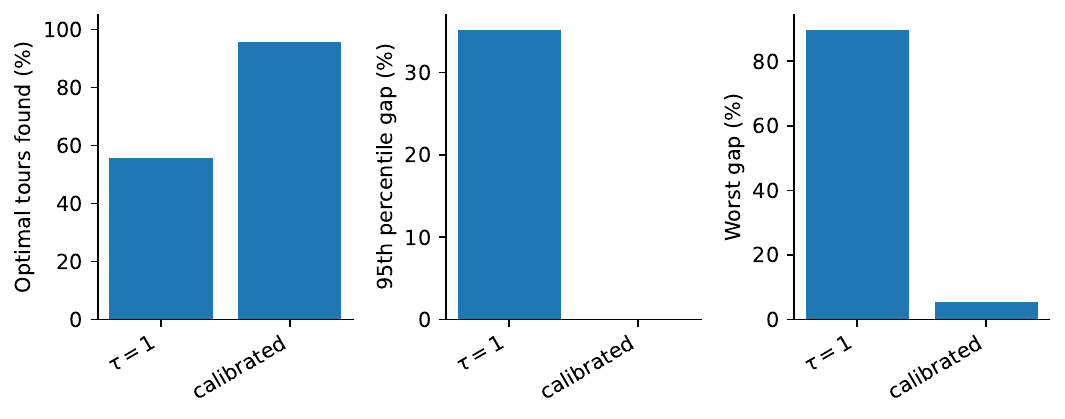}
    \caption{Comparison of $\tau=1$ and the $\tau=\{1,\dots,40\}$ calibration for TN full-filter mode on the small-instance benchmark with known optima. The panels report optimal-solution rate, 95th-percentile relative gap, and worst relative gap.}
    \label{fig:tau-calibration-quality}
  \end{figure}

After selecting the calibrated $\tau$ schedule, we evaluate how much this choice changes TN full-filter mode on the disjoint small-instance evaluation subset with known optima; the comparison is summarized in Fig.~\ref{fig:tau-calibration-quality}. This is no longer a parameter-search step. It compares the original $\tau=1$ baseline and the $\tau=\{1,\dots,40\}$ calibration on the evaluation split.

With $\tau=1$, TN full-filter mode is optimal in 55.56\% of the 90 small-instance evaluations with known optima. The $\tau=\{1,\dots,40\}$ calibration raises this to 95.56\%. With the calibration, valid-route and finite-edge rates are 100\%, the median gap is 0, the 95th-percentile gap is approximately 0, the mean gap is 0.104\%, and the worst gap is 5.56\%. The only four non-optimal cases in this small-instance evaluation are valid finite tours from the random symmetric family: $(n,seed)=(7,8)$ with gap 5.56\%, $(7,9)$ with gap 3.41\%, $(8,8)$ with gap 0.35\%, and $(7,5)$ with gap 0.07\%. The calibration substantially improves optimum recovery on these small synthetic instances, but it does not remove the numerical sensitivity to scale, $\tau$, and finite precision. Thus TN full-filter mode should not be described as always recovering the optimum in finite-precision arithmetic.

\subsection{\texorpdfstring{Solver comparison}{Solver comparison}}\label{ssec:solver-comparison}

\begin{figure}[t]
  \centering
  \includegraphics[width=\linewidth]{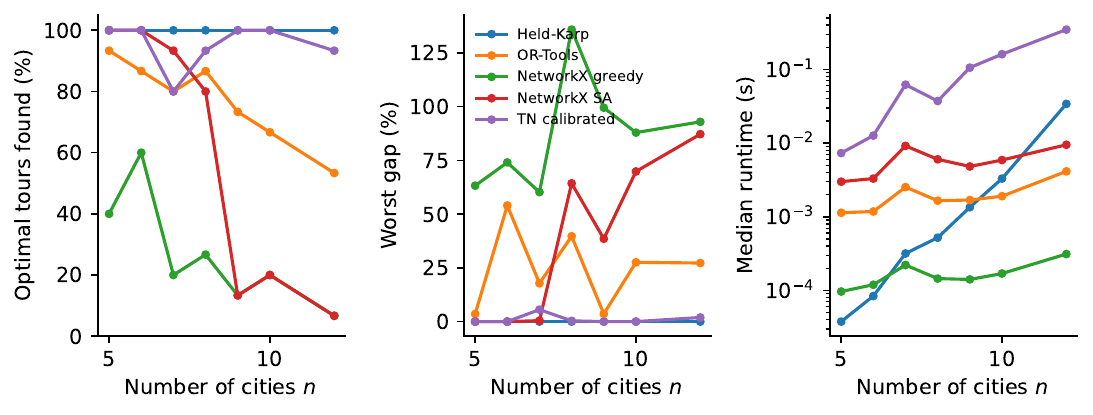}
  \caption{Per-size comparison of Held-Karp, OR-Tools, NetworkX greedy, NetworkX simulated annealing, and TN full-filter mode with the final calibrated $\tau$ schedule on instances with known optima. The panels report optimal-solution rate, worst relative gap, and median runtime; runtime is shown only as an implementation diagnostic.}
  \label{fig:solver-comparison}
\end{figure}

We next place TN full-filter mode with the final calibrated $\tau$ schedule next to representative references: Held-Karp as the exact dynamic-programming reference, OR-Tools~\cite{ortools} as a practical solver wrapper, and NetworkX greedy and simulated annealing heuristics~\cite{networkx}. This comparison, shown in Fig.~\ref{fig:solver-comparison}, is meant to contextualize solution quality on the 105 evaluation instances, not to tune or exhaustively rank classical solvers. The OR-Tools and NetworkX entries should be read under the configurations used here. Aggregated over the benchmark, Held-Karp provides the exact reference and has 100\% optimality. TN full-filter mode provides 95.24\% optimality under the $\tau=\{1,\dots,40\}$ calibration, with median gap 0, mean gap 0.108\%, and worst gap 5.56\%. OR-Tools reaches 77.14\% optimality with the configuration used here, NetworkX greedy reaches 26.67\%, and NetworkX simulated annealing reaches 59.05\%.

The comparison shows that the calibrated TN implementation often returns optimal tours or very small gaps on these small instances, while still falling short of the exact Held-Karp reference in finite precision. The runtime panel is included for reproducibility and implementation context only; these Python implementations and wrappers are not an optimized performance study. NetworkX Christofides is only applicable to Euclidean or symmetric metric instances and is therefore not used as a whole-benchmark aggregate.

\subsection{\texorpdfstring{Restriction-layer ablation}{Restriction-layer ablation}}\label{ssec:layer-ablation}

\begin{figure}[t]
  \centering
  \includegraphics[width=\linewidth]{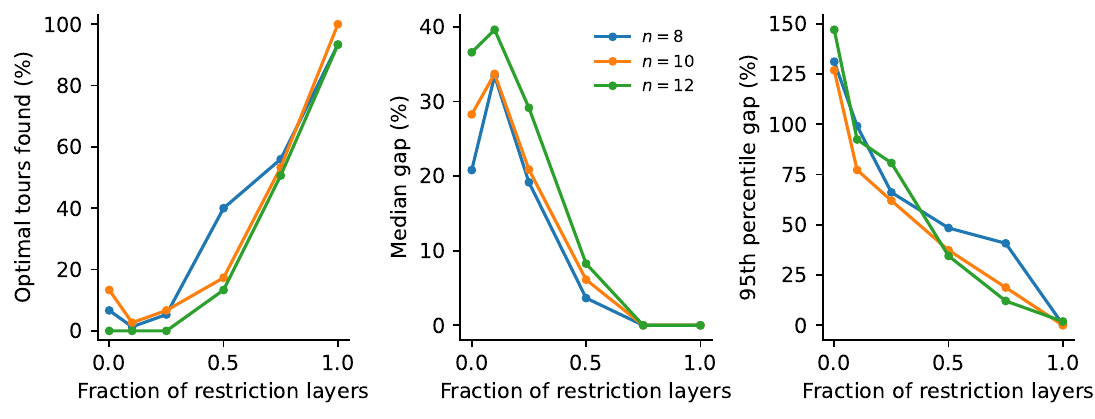}
  \caption{Restriction-layer ablation for the TN solver at $n=8,10,12$. The horizontal axis is the retained fraction of restriction layers; $\rho=1$ is full-filter mode. The panels report optimal-solution rate, median relative gap, and 95th-percentile relative gap.}
  \label{fig:layer-ablation}
\end{figure}

The restriction layers are the mechanism that enforces the non-repetition structure of TSP tours. We isolate the corresponding quality loss with a layer-ablation experiment for $n=\{8,10,12\}$, summarized in Fig.~\ref{fig:layer-ablation}. The tested retained-layer ratios are $\rho\in\{0,0.1,0.25,0.5,0.75,1\}$; $\rho=1$ corresponds to the full-filter mode of the solver.
Each $(n,\rho)$ group contains 75 solver evaluations, obtained from three instance families, five evaluation seeds, and five repeats per setting.

\begin{table}[t]
    \centering
    \caption{Aggregated layer-ablation results across $n\in\{8,10,12\}$, the three instance families, five evaluation seeds, and five repeats.}
    \footnotesize
    \begin{tabular}{rrrr}
        \toprule
        retained ratio $\rho$ & optimal rate & median gap & 95th-percentile gap \\
        \midrule
        0.00 & 6.67\% & 30.45\% & 129.91\% \\
        0.10 & 1.33\% & 36.17\% & 89.67\% \\
        0.25 & 4.00\% & 23.78\% & 66.54\% \\
        0.50 & 23.56\% & 5.47\% & 43.39\% \\
        0.75 & 53.33\% & 0.00\% & 24.36\% \\
        1.00 & 95.56\% & 0.00\% & 0.00\% \\
        \bottomrule
    \end{tabular}
    \label{tab:layer-ablation}
\end{table}
The ablation confirms that these layers are doing substantive constraint-enforcement work. Valid-route and finite-edge rates are 100\% for every retained-layer ratio. The aggregated results in Table~\ref{tab:layer-ablation} show that increasing the retained fraction tends to improve quality, with the strongest result at $\rho=1$. Full-filter/full-layer mode provides 95.56\% under the final $\tau=\{1,\dots,40\}$ calibration. However, the reduced-layer heuristics do not improve uniformly under the wider calibration, and they should not be presented as guaranteeing optimality. The wider calibration mainly improves the full-layer mode.

\subsection{\texorpdfstring{Limitations and interpretation}{Limitations and interpretation}}\label{ssec:benchmark-limitations}

These benchmarks validate correctness and numerical behavior on small synthetic TSP instances; they do not claim computational advantage over specialized classical solvers. The calibration and evaluation seeds are disjoint, but both splits come from the same synthetic families, so the results establish in-distribution behavior on the tested families rather than out-of-distribution generalization. The supported conclusions are narrower: a single fixed $\tau$ is inadequate for this implementation, the wider size-dependent calibration substantially improves solution quality, TN full-filter mode with calibrated $\tau$ often recovers the optimum but not always in finite precision, and restriction layers have a strong empirical effect on solution quality. The main residual limitations are the exponential exact contraction, sensitivity to $\tau$, finite-precision underflow or loss of contrast, degeneracy or near-degeneracy among competing tours, and the limited size and diversity of the test suite.

\subsection{\texorpdfstring{Industrial JRP comparison}{Industrial JRP comparison}}\label{ssec:jrp-comparison}

As a separate industrial integration check, for the productive resolution of the JRP we compared three different methods: D-Wave's \texttt{Advantage\_system4.1} quantum annealer, Azure's Digital Annealer (DA) and the Tensor Networks method explained in this paper. 

Initial comparisons showed good results for the three platforms, but \texttt{Advantage\_system4.1} was discarded due to lack of access to the production scale. 

For the comparison between DA and TN methods, we selected 72 days at random from a whole year, in order to minimize the variations caused by seasonal effects. The day selection was also equiprobably selected amongst the seven days of the week. 

The first thing we did to compare the two methods was to sum the total gain in job quality $\Delta P\equiv \sum_i P^C_i-P^V_{x_i}$ for all days, that is, the accumulated difference between the qualities of the originally assigned jobs and the qualities of the reassigned jobs that each method suggested. It turned out that, for each of the 72 days, $\Delta P_{DA}=\Delta P_{TN}$. Then, since the problem itself minimizes the total cost function $C$ and not only $\Delta P$, we compared the other part of the cost function, $\Delta A\equiv \sum_i A^C_{i,i}-A^V_{x_i,i}$, to break the tie. It did not. However, the solutions were not equivalent because of the existence of degeneracy. For example, we found a case where the only available worker $i$, assigned to a low-quality job, could be reassigned to two different workplaces of the same high quality $P^V_{x_i}$. The worker also had the same affinity $A^V_{x_i,i}$ with the two vacant workplaces. Then, the gain of choosing one or the other was identical, and the two methods -DA and TN- chose at random different non-equivalent answers.

 The agreement between the two methods, with the same accumulated cost over all selected days and differences only in degenerate cases, is a useful consistency check for the integration.

This industrial comparison is separate from the small-instance TSP calibration study above and is intended as an integration validation, not as evidence of a computational advantage over assignment-specific classical methods.

\FloatBarrier
\section{Conclusions}
We have studied a tensor-network formulation for a set of different versions of the Traveling Salesman Problem, characterized the small-instance TSP finite-$\tau$ implementation through explicit $\tau$ calibration and restriction-layer ablation, and have shown its integration in a representative industrial JRP case. Tables~\ref{tab:comparative} and~\ref{tab:comparative_sparse} summarize, respectively, the dense layer-wise and the sparse-local contraction complexities of the main cases presented. In addition, we can combine these tensor-network modules for hybrid cases, such as a group-constrained TSP problem with precedence and non-Markovian costs.

This framework has several important limitations. The first one is exponential scaling to obtain the zero-temperature exact-arithmetic solution rule. Even with the application of the above approximate techniques, to obtain a good result, we need to apply several rules, increasing the number of rules needed with the size of the problem. Regarding compression techniques, due to the projection layers, we need an exponentially large bond dimension to maintain the necessary entanglement in the worst case to obtain a good solution. Accordingly, the exact tensor-network formulation presented here should be understood as a formal exact-arithmetic exponential-time representation and as a platform for controlled approximations and extensions, rather than as a polynomial-time algorithm for the general TSP.

Therefore, although both methods allow for dealing with larger cases, depending on the instances to be solved, they may still require exponential memory and time.

The second constraint is numerical stability. That is, we have to balance the number of states we are summing when tracing, since we are summing an exponential number of amplitudes, and the damping factor $\tau$, since it has to be large to distinguish the maximum, but not so large as to make all amplitudes go to zero. Degenerate or nearly degenerate optima are especially delicate, because finite precision may make several branches numerically indistinguishable even when the zero-temperature exact-arithmetic rule is well defined. The sparse-local contractions may be evaluated either in amplitude space or in log-space. The latter is preferable for large $\tau$ or large instances, replacing sums of Boltzmann weights by log-sum-exp operations.

The experiments above already show that finite-$\tau$ and finite-precision effects can change the selected tour, and these effects may become more severe for larger or more weakly separated instances. It could be mitigated by partitioning the problems, applying state rescaling or log-sum-exp techniques, or simply allowing the states with higher cost to disappear due to the numerical error, since we really only need to maintain one amplitude, that of the valid state with lowest cost.

The deterministic sparsity of the local $F$, $S$, and $Z$ tensors can be exploited without changing the tensor network or the contraction scheme, reducing polynomial time prefactors but not the exponential factor. If the intermediate tensors are still stored densely, the asymptotic memory bounds remain those of the dense analysis; memory reductions require sparse storage of the intermediate tensors and a separate support count. In all cases, the exact-arithmetic full-constraint method remains exponential in the worst case. Layer removal should therefore be read as a heuristic approximation, and MPS compression as an approximate representation whose quality depends on the instance and retained bond dimension. The experiments in this paper are consequently illustrative small-instance studies and an industrial integration check, not evidence of broad dominance over specialized TSP, assignment, or min-cost-flow solvers.

\begin{table*}
    \centering
    \caption{Dense layer-wise contraction complexity with and without reuse of intermediate calculations.}
    \begingroup
    \footnotesize
    \setlength{\tabcolsep}{3pt}
    \renewcommand{\arraystretch}{1.18}
    \begin{tabularx}{\textwidth}{@{}lYYYY@{}}
        \toprule
        Case & Steps without reuse & Space without reuse & Steps with reuse & Space with reuse\\
        \midrule
        TSP & $\hat{N}^4 2^{\hat{N}}$ & $\hat{N}^2 2^{\hat{N}}$ & $\hat{N}^4 2^{\hat{N}}$ & $\hat{N}^3 2^{\hat{N}}$\\
        DNSNN & $N^3N_s^2D^N$ & $N^2D^N$ & $N^3N_sD^N$ & $N^2N_sD^N$\\
        NMTSP & $N^{2K+1}N_s^2D^N$ & $N^{K+1}D^N$ & $N^{2K+1}N_sD^N$ & $N^{K+1}D^N\max\left(N,N_s\right)$\\
        BTSP & $\hat{N}M_c 2^{\hat{N}}\max\left(\hat{N}, M_c \right)$ & $\hat{N}M_c 2^{\hat{N}}$ & $\hat{N}M_c 2^{\hat{N}}\max\left(\hat{N}, M_c \right)$ & $\hat{N}^2M_c 2^{\hat{N}}$\\
        PTSP & $N^3 2^M$ & $N^2 2^M$ & $N^3 2^M$ & $N^3 2^M$\\
        TSPP & $\hat{N}^4 2^{\hat{N}}$ & $\hat{N}^2 2^{\hat{N}}$ & $\hat{N}^4 2^{\hat{N}}$ & $\hat{N}^3 2^{\hat{N}}$\\
        \bottomrule
    \end{tabularx}
    \endgroup
    \label{tab:comparative}
\end{table*}

The sparse-local bounds in Table~\ref{tab:comparative_sparse} refer to arithmetic operations. The memory bounds remain those of the dense intermediate representation unless the intermediate tensors $A^i$, $W^i$, and $V^i$ are also stored sparsely. We do not claim such memory reductions in this table.

\begin{table*}
    \centering
    \caption{Local sparse contraction of the same tensor networks. The time bounds exploit the zero pattern of the local tensors while keeping the same contraction scheme. Memory with dense intermediates matches Table~\ref{tab:comparative}.}
    \begingroup
    \footnotesize
    \setlength{\tabcolsep}{3pt}
    \renewcommand{\arraystretch}{1.18}
    \begin{tabularx}{\textwidth}{@{}lYY>{\raggedright\arraybackslash}X@{}}
        \toprule
        Case & Dense time & Sparse-local time & Comment\\
        \midrule
        TSP & $\mathcal{O}\left(\hat{N}^42^{\hat{N}}\right)$ & $\mathcal{O}\left(\hat{N}^32^{\hat{N}}\right)$ & sparse $F$ and $S$ contractions remove one polynomial factor\\
        DNSNN & $\mathcal{O}\left(N^3N_s^2D^N\right)$ & $\mathcal{O}\left(N^2N_s^2D^N\right)$ & $F$-counter transition has $\mathcal{O}(ND)$ nonzeros\\
        DNSNN with reuse & $\mathcal{O}\left(N^3N_sD^N\right)$ & $\mathcal{O}\left(N^2N_sD^N\right)$ & same local counter sparsity\\
        BTSP & $\mathcal{O}\left(\hat{N}M_c2^{\hat{N}}\max(\hat{N},M_c)\right)$ & $\mathcal{O}\left(\hat{N}^2M_c2^{\hat{N}}\right)$ & bottleneck update $p=\max(q,E_{k,i})$ is deterministic\\
        PTSP & $\mathcal{O}\left(N^32^M\right)$ & $\mathcal{O}\left(N^22^M\right)$ & class filters are sparse binary projectors\\
        JRP, $I\le J$ & $\mathcal{O}\left(J^2I^22^I\right)$ & $\mathcal{O}\left(J^2I2^I\right)$ & vacancy filters remove one local factor $I$\\
        JRP with reuse, $I\le J$ & $\mathcal{O}\left(JI^22^I\right)$ & $\mathcal{O}\left(JI2^I\right)$ & no $S$ layer; dummy state has no filter\\
        \bottomrule
    \end{tabularx}
    \endgroup
    \label{tab:comparative_sparse}
\end{table*}

For the JRP rows in Table~\ref{tab:comparative_sparse}, if $I>J$, swap $I$ and $J$, as in Sec.~\ref{sec:once}.

Future lines of research could include an optimized implementation of sparse-local contractions, systematic experiments comparing dense and sparse TN kernels, combinations with MPS compression, and a deeper analysis of layer-elimination heuristics.

\bibliographystyle{unsrt}  
\bibliography{references}

\end{document}